\documentclass[11pt, onecolumn]{article}
\usepackage[utf8]{inputenc}
\usepackage{helvet}
\usepackage{amsfonts}
\usepackage{amsmath}
\usepackage{amssymb}
\usepackage{tocloft}
\usepackage{url}

\DeclareFontFamily{U}{mathb}{\hyphenchar\font45}
\DeclareFontShape{U}{mathb}{m}{n}{
      <5> <6> <7> <8> <9> <10> gen * mathb
      <10.95> mathb10 <12> <14.4> <17.28> <20.74> <24.88> mathb12
      }{}
\DeclareSymbolFont{mathb}{U}{mathb}{m}{n}
\DeclareFontSubstitution{U}{mathb}{m}{n}

\let\dot\relax
\DeclareMathAccent{\dot}{0}{mathb}{"39}
\let\ddot\relax
\DeclareMathAccent{\ddot}{0}{mathb}{"3A}
\let\dddot\relax
\DeclareMathAccent{\dddot}{0}{mathb}{"3B}
\let\ddddot\relax
\DeclareMathAccent{\ddddot}{0}{mathb}{"3C}

\usepackage{graphicx}

\newcommand{\ohwidth}{0.75\columnwidth}
\newcommand{\swidth}{0.5\columnwidth}

\newcommand{\mwidth}{0.6\columnwidth}

\newcommand{\T}{\rule{0pt}{2.6ex}}

\usepackage{threeparttable}
\usepackage{multirow}
\usepackage{setspace}
\usepackage{gensymb}

\usepackage{color}
\usepackage[nomarkers,tablesfirst, nolists]{endfloat}
\usepackage{float}
\newfloat{supptable}{htbp}{lost}

\floatname{supptable}{Supporting Table}%
\DeclareDelayedFloat{supptable}{Supporting Tables}
\newfloat{suppfigure}{htbp}{losf}

\floatname{suppfigure}{Supporting Figures and Tables}%
\DeclareDelayedFloat{suppfigure}{Supporting Figures}
\usepackage{comment}
\newif\ifsupp
\supptrue

\usepackage{cite}
\let\citeleft=(
\let\citeright=)
\usepackage{cleveref}
\crefname{equation}{eq.}{eqs.}
\crefname{figure}{fig.}{figs.}
\crefname{table}{table}{tables}

\begin{document}
\bibliographystyle{mrm}

\begin{titlepage}

\begin{center}
	\begin{Large}
		\begin{bf}
Reconstruction by Calibration over Tensors for Multi-Coil Multi-Acquisition Balanced SSFP Imaging
		\end{bf}
	\end{Large}
\end{center}
\bigskip
\begin{center}
Erdem Biyik$^{1,2}$, Efe Ilicak$^{1,2}$, Tolga \c{C}ukur$^{1,2,3}$
\end{center}
\vspace*{0.1in}
\noindent
$^1$Department of Electrical and Electronics Engineering, Bilkent University, Ankara, Turkey\\
$^2$National Magnetic Resonance Research Center (UMRAM), Bilkent University, Ankara, Turkey\\
$^3$Neuroscience Program, Sabuncu Brain Research Center, Bilkent University, Ankara, Turkey

\vspace*{0.1in}
\noindent
{\em Running title:} Reconstruction by Calibration over Tensors for Multi-Coil Multi-Acquisition bSSFP\\
                                                           
\noindent
{\em Address correspondence to:} \\
	Tolga \c{C}ukur \\
	Department of Electrical and Electronics Engineering, Room 304 \\
	Bilkent University \\
	Ankara, TR-06800, Turkey \\
	TEL: +90 (312) 290-1164 \\
    E-MAIL: cukur@ee.bilkent.edu.tr

\vspace*{0.2in}

\noindent
This work was supported in part by a Marie Curie Actions Career Integration Grant (PCIG13-GA-2013-618101), by a European Molecular Biology Organization Installation Grant (IG 3028), by the Turkish Academy of Sciences TUBA GEBIP program, and the Science Academy BAGEP award.

\vspace*{0.2in}

\noindent
Submitted to {\it Magnetic Resonance in Medicine}.\\
Final published version: \url{http://onlinelibrary.wiley.com/doi/10.1002/mrm.26902/abstract}.\\
\end{titlepage}

\section*{Abstract}
\setlength{\parindent}{0in}
\textbf{Purpose:} To develop a rapid imaging framework for balanced steady-state free precession (bSSFP) that jointly reconstructs undersampled data (by a factor of R) across multiple coils (D) and multiple acquisitions (N). To devise a multi-acquisition coil compression technique for improved computational efficiency. \\
\textbf{Methods:} The bSSFP image for a given coil and acquisition is modeled to be modulated by a coil sensitivity and a bSSFP profile. The proposed reconstruction by calibration over tensors (ReCat) recovers missing data by tensor interpolation over the coil and acquisition dimensions. Coil compression is achieved using a new method based on multilinear singular value decomposition (MLCC). ReCat is compared with iterative self-consistent parallel imaging (SPIRiT) and profile encoding (PE-SSFP) reconstructions. \\
\textbf{Results:} Compared to parallel imaging or profile-encoding methods, ReCat attains sensitive depiction of high-spatial-frequency information even at higher R. In the brain, ReCat improves peak SNR (PSNR) by 1.1$\pm$1.0 dB over SPIRiT and by 0.9$\pm$0.3 dB over PE-SSFP (mean$\pm$std across subjects; average for N=2-8, R=8-16). Furthermore, reconstructions based on MLCC achieve 0.8$\pm$0.6 dB higher PSNR compared to those based on geometric coil compression (GCC) (average for N=2-8, R=4-16). \\
\textbf{Conclusion:} ReCat is a promising acceleration framework for banding-artifact-free bSSFP imaging with high image quality; and MLCC offers improved computational efficiency for tensor-based reconstructions. \\

\vspace{0.5in}
\setlength{\parindent}{0in}
{\bf Keywords: bSSFP, accelerated MRI, joint reconstruction, tensor, encoding, coil compression}

\clearpage

\section*{Introduction}
Balanced SSFP sequences are commonly employed in rapid imaging due to their relatively high signal efficiency \cite{SchefflerEur}. While the speed advantage can be countered in part by the $T_2/T_1$ contrast and system imperfections \cite{Bieri05,Bangerter04}, multiple phase-cycled acquisitions can enable improvements in tissue contrast through fat-water separation \cite{Hargreaves:2006bn, ATRDIXON, Cukur:2009do} and in reliability against field inhomogeneity \cite{Bangerter04,Cukur:2007dx,Elliott07,Quist:2012kx}. Yet acceleration techniques are needed to maintain scan efficiency with higher number of acquisitions (N). 

Several approaches were recently proposed for accelerating phase-cycled bSSFP imaging. One study used simultaneous multi-slice imaging on each acquisition \cite{Wang:2015hh}. Undersampled data were recovered via parallel-imaging (PI) reconstructions \cite{PruessmannSENSE,GriswoldGRAPPA} across multiple coils to achieve modest acceleration factors (R $\approx$ 2-3). In \cite{Cukur:2015ic}, we used disjoint variable-density sampling patterns across phase cycles at similar R $\approx$ 4. Independent compressed-sensing (CS) reconstructions \cite{Block,MikiCS,CSENS} were then performed on each acquisition. To further enhance the image quality by improving the kernel estimation, we more recently proposed a profile-encoding framework (PE-SSFP) to jointly reconstruct data from separate phase-cycles \cite{Ilicak:FJpKoYYb}. PE-SSFP yielded improved preservation of high-spatial-frequency details at relatively high R $\approx$ 6-8 compared to conventional PI and CS reconstructions. These previous approaches leverage only a subset of correlated structural information, either across multiple coils or across multiple acquisitions. However, recent studies indicate that joint processing of coils and acquisitions can improve performance for heavily undersampled datasets \cite{Majumdar:2011hj,Bilgic:2011jv,Jin:2016cz}.

\begin{figure}[t]
	\begin{center}
		\includegraphics[width=\mwidth]{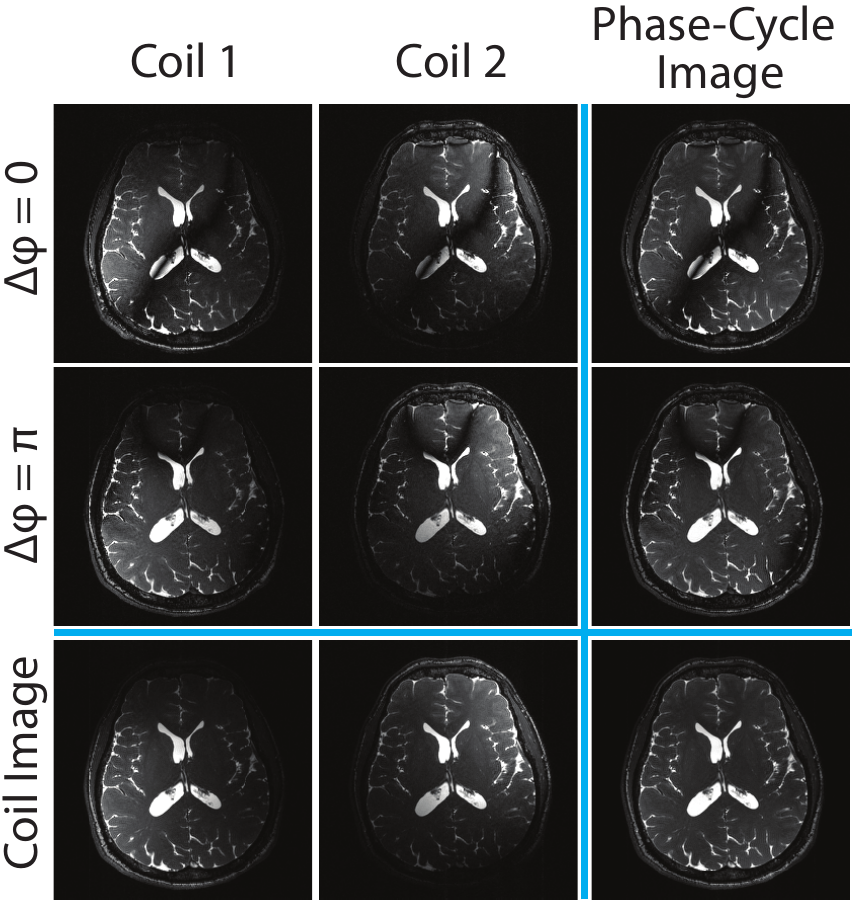}
		\caption{Balanced SSFP images from two phase-cycled acquisitions and two coils are shown. The first two rows show the acquisitions, with $\Delta \phi$ denoting the phase-cycling increment. Similarly, the first two columns show the coils. Acquisition-combined coil images (third row) and coil-combined phase-cycle (third column) images are also shown along with the reference image combined across both coils and acquisitions. Intensity modulations due to bSSFP profiles differ across acquisitions, whereas those due to coil sensitivities vary across coils.	
		}
		\label{fig:acs_coils}
	\end{center}
\end{figure}

Here, we propose an improved framework for phase-cycled bSSFP imaging, reconstruction by calibration over tensors (ReCat), that utilizes correlated information simultaneously across multiple coils and acquisitions (Fig.~\ref{fig:acs_coils}). ReCat is based on a joint encoding model: the bSSFP image for a given coil and phase-cycle is taken to be spatially modulated by a respective pair of coil sensitivity \cite{PruessmannSENSE, Lustig:2010hs} and bSSFP profile \cite{mikisFOV,Ilicak:FJpKoYYb}. A tensor-interpolation kernel comprising coil and acquisition dimensions is estimated from calibration data. This kernel is then used to linearly synthesize unacquired samples. Compared to kernels trained only on coil or on phase-cycles, the ReCat kernel aims to optimize use of aggregate information across both dimensions. 

Joint reconstruction of a multi-coil, multi-acquisition dataset poses significant computational burden. Since modern coils contain a large number of elements, a common approach is either hardware- \cite{king2010optimum} or software-based \cite{buehrer2007array,huang2008software} coil compression. A recent technique is the data-driven geometric coil compression (GCC) that accounts for spatially-varying coil sensitivities across three-dimensional (3D) datasets \cite{Zhang:2013df}. While software-based methods such as GCC can estimate virtual coils separately for each bSSFP acquisition, they ignore shared information about coil sensitivities across acquisitions, yielding suboptimal estimates. Furthermore, the virtual-coil sensitivities in separate acquisitions can be inconsistent due to variations in bSSFP profiles and noise. These limitations can in turn degrade the quality of joint reconstructions. 

To address these limitations, here we propose a new multilinear coil compression (MLCC) technique based on multilinear singular value decomposition for multi-coil, multi-acquisition datasets. It performs compression via tensor-based separation of the coil and acquisition dimensions. It therefore leverages shared coil-sensitivity information to produce a consistent set of virtual coils across acquisitions. 

Comprehensive simulation and in vivo results are presented to demonstrate the potential of the proposed framework for accelerated bSSFP imaging. ReCat significantly improves image quality over both PI reconstruction of multi-coil and CS reconstruction of multi-acquisition data. In addition, reconstructions based on MLCC show superior quality compared to those based on GCC.

\clearpage
\section*{Methods}
The main aim of this study is to enable highly accelerated phase-cycled bSSFP imaging via an expanded framework (ReCat) that jointly processes data aggregated across multiple coils and acquisitions. We start this section with an overview of accelerated bSSFP imaging, and then describe the reconstruction and coil-compression components of ReCat. 

\subsection*{Accelerated Phase-Cycled bSSFP Imaging}
Phase-cycled bSSFP imaging acquires multiple images with different phase increments in radio-frequency excitations. The bSSFP signal at each spatial location $r$ is given by \cite{Bjork:2014hm}:
\begin{equation}
\label{eq:sind}
S_{n,d}(r)={C_d(r)}{M(r)}\frac{{{e^{i\left( {\phi(r)  + \Delta {\phi _n}} \right)/2}} \left( {1 - A(r){e^{ - i\left( {\phi(r)  + \Delta {\phi _n}} \right)}}} \right)}}{{1 - B(r)\cos (\phi(r)  + \Delta {\phi _n})}}
\end{equation}
under the assumption that the echo time (TE) is one half of the repetition time (TR). Here, $S_{n,d}(r)$ denotes the signal captured by the $d^{th}$ coil element ($d \in$ [1 D]) and the $n^{th}$ acquisition ($n \in$ [1 N]). $C_d(r)$ is the coil sensitivity, $\Delta \phi_n$ is the phase increment, and $\phi (r)$ is the phase accrued due to off-resonance (assumed to be constant across acquisitions). Note that $M$, $A$, $B$ are terms that do not depend on off-resonance or phase increments. With $\Delta \phi_n$ equispaced across $[0 \mbox{ }2\pi) $, banding artifacts in separate bSSFP images will be largely nonoverlapping \cite{Scheffler:2003cu}. Thus multiple phase-cycled bSSFP images can be combined to effectively suppress banding artifacts \cite{Bangerter04,Cukur:2007dx}. However, to maintain scan efficiency, each phase-cycled acquisition should first be undersampled by a factor of R, and images must be recovered during subsequent reconstructions. 

One acceleration approach is to use uniform-density deterministic patterns and perform separate PI reconstructions of multi-coil data for each acquisition \cite{Wang:2015hh}. The PI approach casts on \Cref{eq:sind} an encoding model based on coil sensitivities \cite{PruessmannSENSE}:  
\begin{equation}
{S_{n,d}}(r)={C_d}(r){S_n}(r)
\end{equation}
where $S_n (r)$ denotes the phase-cycled image devoid of coil-sensitivity modulation. Autocalibration is typically used to estimate $C_d (r)$ from fully-sampled central k-space data. Separate linear inverse problems are solved to recover $S_n (r)$, which are then combined across acquisitions.

We recently proposed to use variable-density random patterns and perform profile-encoding (PE) reconstructions of multi-acquisition data for each coil \cite{Ilicak:FJpKoYYb}. This PE approach casts on \Cref{eq:sind} an encoding model based on spatial bSSFP profiles \cite{mikisFOV,Cukur:2008ht}:
\begin{equation}
{S_{n,d}}(r)={P_n}(r){S_d}(r)
\end{equation}
where $S_d (r)$ denotes the coil image devoid of bSSFP-profile modulation. $P_n (r)$ can again be estimated from fully-sampled central k-space data. Separate linear inverse problems are solved to recover $S_d (r)$, which are then combined across coils.

PI leverages correlated structural information across coils, whereas PE-SSFP leverages correlated information across acquisitions. Neither technique aggregates information in these two dimensions. This poses a limitation in the recovery of unacquired data and the achievable acceleration rates.

\subsection*{Reconstruction by Calibration over Tensors}
\begin{figure}[t]
  \begin{center}
    \includegraphics[width=\ohwidth]{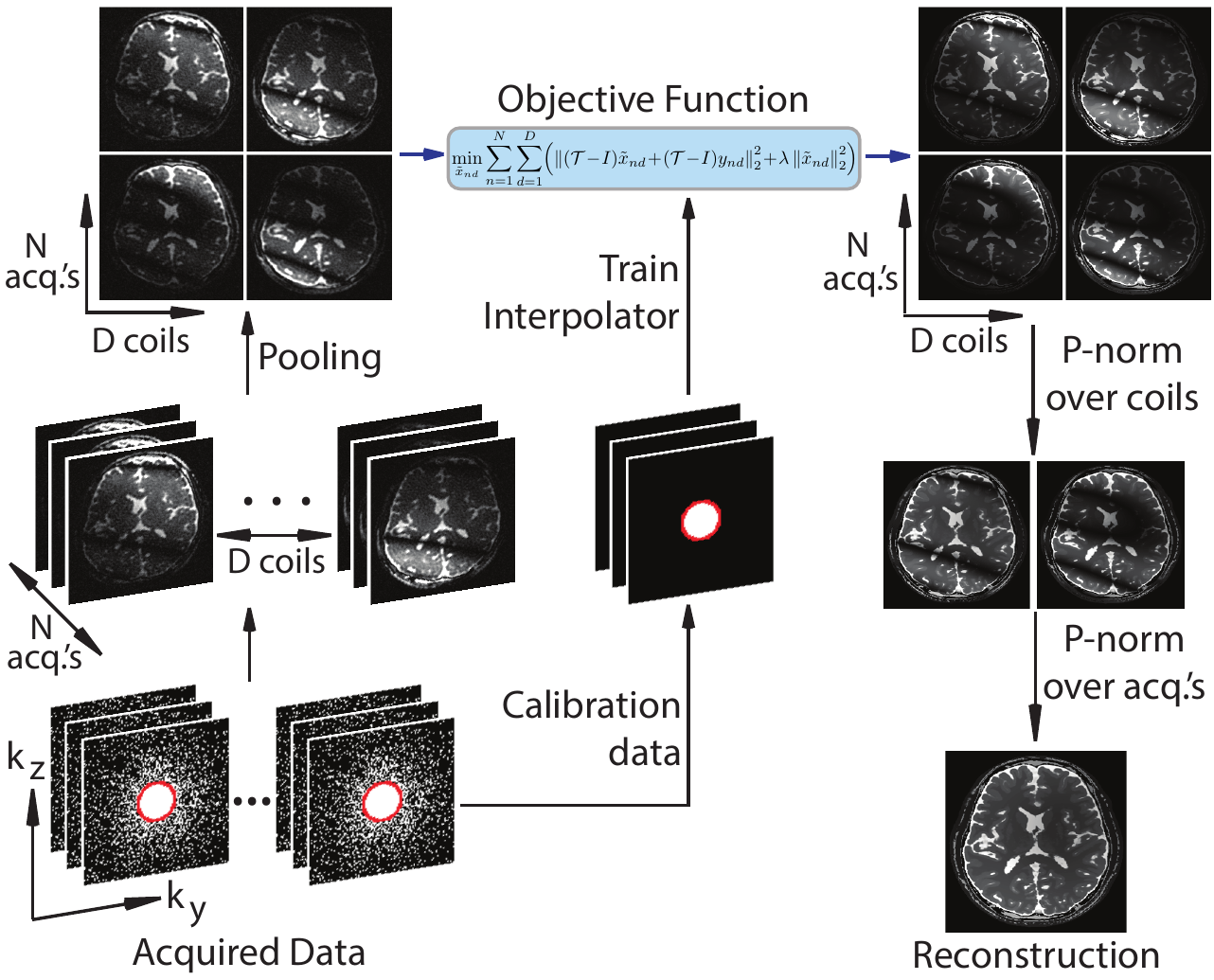}
    \caption{Flowchart of the proposed ReCat method. ReCat reconstructs phase-cycled bSSFP images by jointly processing data from D coils and N acquisitions. An interpolation kernel across coils and acquisitions is estimated from central calibration data. Missing data are iteratively synthesized using this kernel. Reconstructed images are first combined over coils and then over acquisitions with the p-norm method.
    }
    \label{fig:recat}
  \end{center}
\end{figure}

Here we propose to accelerate multi-coil multi-acquisition bSSFP imaging via a new technique named reconstruction by calibration over tensors (ReCat). Unlike PI or PE-SSFP, the proposed approach utilizes correlated information simultaneously across the coil and acquisition dimensions. For this purpose, ReCat casts on \Cref{eq:sind} a tensor encoding model based on both coil sensitivities and bSSFP profiles:
\begin{equation}
\label{eq:invprob}
{S_{n,d}}(r)={P_n}(r){C_d}(r){S_o}(r)
\end{equation}
where $S_o (r)$ denotes the ideal bSSFP image devoid of modulations due to both bSSFP profiles and coil sensitivities. Leveraging this model, ReCat recovers unacquired k-space data in terms of collected data $y_{n,d}$ aggregated across coil and acquisition dimensions (see Fig.~\ref{fig:recat}). First, a tensor-based interpolation kernel is estimated from fully-sampled calibration data. This kernel is then used to linearly synthesize missing k-space samples.

\textit{Interpolation kernel:} ReCat uses an interpolation kernel to estimate each unacquired k-space sample as a weighted combination of neighboring data in all coils and acquisitions:
\begin{equation}
\label{eq:interp}
{x_{n,d}}({k_r})=\sum\limits_{i=1}^N {\sum\limits_{j=1}^D {{t_{ij,nd}}({k_r}) \otimes {x_{i,j}}({k_r})} } 
\end{equation}
where $x_{n,d}$ is the k-space data from the $n^{th}$ acquisition and $d^{th}$ coil, $k_r$ is the k-space location, and $\otimes$ is a convolution. The kernel $t$ is a third-order tensor; and ${t_{ij,nd}}({k_r})$ reflects the linear contribution of samples in $k_r$'s neighborhood from the $i^{th}$ acquisition and $j^{th}$ coil, onto the sample at $k_r$ from the $n^{th}$ acquisition and $d^{th}$ coil. The unknown kernel weights are obtained from calibration data $y^{c}$, a fully-sampled central region of k-space. The calibration procedure finds the weights that are consistent with the calibration data according to \Cref{eq:interp}. This leads to the following least-squares solution:
\begin{equation}
{t_{nd}}={({Y^*}Y + \beta I)^{ - 1}}{Y^*}{y^{c}_{n,d}}
\end{equation}
where ${t_{ij,nd}}$ are concatenated to form $t_{nd}$, and $y^{c}$ are aggragated in matrix form $Y$. The regularization parameter $\beta$ is used to improve matrix conditioning and noise resilience \cite{Murphy:2012hq}. 

In this study, we prescribed an interpolation kernel that covered a 11$\times$11 neighborhood of k-space samples as in \cite{Ilicak:FJpKoYYb}. The regularization weight $\beta$ was varied in the range $(0, 0.2]$. An optimized value of $\beta=0.05$ was determined on simulated phantoms (see Sup. Fig. S1a), and used in all subsequent reconstructions. Finally, the convolution operations in \Cref{eq:interp} were transformed into matrix form for convenience:
\begin{equation}
\label{eq:interp2}
{x}=\mathcal{T} {x} 
\end{equation}
This matrix operator $\mathcal{T}$ was used to linearly synthesize unacquired samples during reconstruction. 

\textit{Reconstruction:} ReCat recovers missing k-space samples based on the interpolation operator $\mathcal{T}$. Inspired by the SPIRiT method (iterative self-consistent parallel imaging reconstruction) for multi-coil imaging \cite{Lustig:2010hs}, a self-consistency formulation is used that enforces consistency of both acquired and recovered data with \Cref{eq:interp2}. Accordingly, ReCat solves the following optimization problem:
\begin{equation}
\label{eq:recat_opt}
\mathop {\min }\limits_{{{\tilde x}_{nd}}} \quad \sum\limits_{n=1}^N {\sum\limits_{d=1}^D {\left( {\left\| {(\mathcal{T} - I){{\tilde x}_{nd}} + (\mathcal{T} - I){y_{nd}}} \right\|_2^2 + \lambda \left\| {{{\tilde x}_{nd}}} \right\|_2^2} \right)} }
\end{equation}
Here ${\tilde x _{nd}}$ denote the unacquired data to be recovered, and $y_{nd}$ denote the acquired data from the n$^{th}$ acquisition and d$^{th}$ coil. The separation of ${\tilde x}$ from $y$ ensures that acquired samples are unchanged during reconstruction. An ${\ell}_2$-regularization term with weight $\lambda$ is used to penalize the energy in recovered k-space samples. 

In this study, the unconstrained optimization in \Cref{eq:recat_opt} was expressed as a linear system of equations, and solved using the iterative least squares (LSQR) method. A total of 20 iterations were sufficient to obtain stable reconstructions. The regularization weight $\lambda$ was varied in the range $(0, 0.03]$. An optimized value of $\lambda=0.018$ was determined on simulated phantoms (see Sup. Fig. S1b), and used in all subsequent reconstructions. This value was observed to yield a good compromise between suppression of aliasing interference and preservation of structural details.

To demonstrate ReCat, zero-filled Fourier (ZF), SPIRiT \cite{Lustig:2010hs} and PE-SSFP \cite{Ilicak:FJpKoYYb} reconstructions were also implemented. In ZF, zero-filled k-space data were compensated for variable sampling density and inverse Fourier transformed to obtain images for each acquisition and each coil. In SPIRiT, multi-coil data from each acquisition were independently reconstructed by removing the coil dimension from \Cref{eq:recat_opt}. In PE-SSFP, multi-acquisition data from each coil were independently reconstructed by removing the acquisition dimension from \Cref{eq:recat_opt}. Both SPIRiT and PE-SSFP reconstructions were obtained via the LSQR algorithm with 20 iterations and identical $\beta$, $\lambda$ to ReCat.

All reconstruction methods produced separate images from each acquisition and each coil. Individual images were then combined with the p-norm method to maintain favorable performance in artifact suppression and SNR efficiency \cite{PNORM}. Combination was performed with $p=2$ across coils, and with $p=4$ across acquisitions (see Sup. Fig. S2). Two different orders of combination were tested: first across coils then acquisitions, and first across acquisition then coils. No significant difference was observed due to combination order.

All reconstruction algorithms were executed in MATLAB (MathWorks, MA). The implementations used libraries in the SPIRiT toolbox \cite{Lustig:2010hs}. The ReCat algorithm is available for general use at: \url{http://github.com/icon-lab/mrirecon}.

\subsection*{Multi-Linear Coil Compression}
As $N$ and $D$ grow, it becomes demanding to compute the interpolation kernel $\mathcal{T}$ and to jointly reconstruct multi-coil multi-acquisition datasets. To improve computational efficiency, coil-compression techniques are typically employed to map $D$ coils onto $D'$ virtual coils \cite{buehrer2007array,huang2008software,king2010optimum,Zhang:2013df}. Hardware-based compression is suboptimal since it ignores variability in coil sensitivity due to subject configuration \cite{king2010optimum}. Meanwhile, conventional software-based methods either rely on explicit knowledge of coil sensitivities  \cite{buehrer2007array} or assume spatially-invariant sensitivities across the imaging volume \cite{huang2008software}. 

To alleviate these limitations, a geometric coil compression (GCC) was recently proposed for single-acquisition 3D Cartesian imaging that performs data-driven compression separately for each spatial location in the readout dimension \cite{Zhang:2013df}. While GCC can cope with spatially-varying coil sensitivities, it disregards shared sensitivity information across acquisitions. Furthermore, since GCC is performed independently on each acquisition, the resulting virtual coils can have inconsistent spatial sensitivities across acquisitions. As a result, accuracy of virtual-coil estimates can be impaired in the presence of noise, and joint reconstructions can be suboptimal due to coil inconsistency. 

Here we propose a new method called multi-linear coil compression (MLCC) for Cartesian sampling based on multi-linear singular value decomposition (SVD). MLCC performs joint compression of multi-slice, multi-coil, multi-acquisition bSSFP data. Therefore, it identifies a shared set of virtual coils across acquisitions, as opposed to GCC that identifies independent sets of coils for separate acquisitions. Note that, when disjoint sampling patterns are prescribed, unacquired locations differ among acquisitions. A simple compression of data pooled across coils and acquisitions would produce nonzero data in many unacquired locations, leading to substantial information loss during reconstruction. Instead, MLCC first models bSSFP data as a fifth-order tensor $\mathcal{A}$ of size $\mathbb{I}_{r1}\!\times\!\mathbb{I}_{r2}\!\times\!\mathbb{I}_{r3}\!\times\!N\!\times\!D$, where $\mathbb{I}_{r1,2}$ denote data size in two phase-encode dimensions, and $\mathbb{I}_{r3}$ is the number of cross-sections in the readout dimension. MLCC then approximates this tensor with reduced size in the coil dimension $D'$.

Tensor theory indicates that any complex tensor of order $H$ can be expressed as the product of a core tensor with unitary matrices in each dimension \cite{de2000multilinear}:
\begin{equation}
\label{eq:mlsvd_theorem}
\mathcal{A}=\mathcal{S}\!\times_1\!\mathbf{U}^{(1)}\!\times_2\!\mathbf{U}^{(2)}\dots\!\times_H\!\mathbf{U}^{(H)}
\end{equation}
where $\mathcal{S}$ is the \textit{core tensor} of size $\mathbb{I}_{1}\!\times\mathbb{I}_2\!\times\dots\times\mathbb{I}_H\!$, $\mathbf{U}^{(h)}$ is a unitary ($\mathbb{I}_h\!\times\mathbb{I}_h$)-matrix, and $\times_h$ denotes the $h$-mode tensor-matrix product. This multi-linear SVD calculates the core tensor, unitary matrices in each dimension, along with n-mode singular values $\sigma_i^{(h)}$ ($\sigma_1^{(h)} \geqslant \sigma_2^{(h)} \geqslant \dots \geqslant \sigma_H^{(h)} \geqslant 0$). The tensor can then be decomposed along dimension $h$ by constructing a set of $\mathbb{I}_h$ subtensors $\mathcal{A}'$ along mode-$h$:
\begin{eqnarray}
 \mathcal{A} &=& (\mathcal{S}\!\times_1\!\mathbf{U}^{(1)}\dots\!\times_H\!\mathbf{U}^{(H)}) \!\times_h\!\mathbf{U}^{(h)} \\ \nonumber
&=& \mathcal{A}' \!\times_h\!\mathbf{U}^{(h)} 
\end{eqnarray} 

In MLCC, a fifth-order tensor $\mathcal{A}$ is formed from undersampled data across all coils and acquisitions. This tensor is then decomposed along the coil (fifth) dimension via multi-linear SVD:
\begin{eqnarray}
\label{eq:new_tensor}
& \mathcal{A}'=\mathcal{S}\!\times_1\!\mathbf{U}^{(1)}\!\times_2\!\mathbf{U}^{(2)}\!\times_3\!\mathbf{U}^{(3)}\!\times_4\!\mathbf{U}^{(4)}
\end{eqnarray}
where a set of $D$-many coil subtensors is obtained $\mathcal{A}'=\{\mathcal{A}'_i, i \in [1, D] \}$ with individual subtensors ordered according to the coil-mode singular values. As such, data can be mapped onto $D'$ virtual coils by retaining the first $D'$ subtensors $\{\mathcal{A}'_1, \mathcal{A}'_2, \dots, \mathcal{A}'_{D'} \}$ that account for the highest amount of variance in the data. Note that the unitary matrix in the coil dimension satisfies:
\begin{equation}
\label{eq:mlsvd_accelerate2}
\mathbf{U}^{(5)^T} \times \mathbf{U}_{1:D,1:D'}^{{(5)}}=\begin{bmatrix}
	\mathbf{I}_{D' \times D'} & \mathbf{0}_{(D\!-\!D')\times D'}
 \end{bmatrix}^T 
\end{equation}
where $\mathbf{I}$ is the identity and $\mathbf{0}$ is the zero matrix. The tensor approximation can then be expressed as:
\begin{eqnarray}
\label{eq:mlsvd_accelerate3}
\hat {\mathcal{A}} &=& \mathcal{A}'\times_5 \mathbf{U}^{(5)}\times_5\mathbf{U}_{1:D,1:D'}^{{(5)}^T} \\
&=& \mathcal{A}\times_5 \mathbf{U}_{1:D,1:D'}^{{(5)}^T} \nonumber
\end{eqnarray}
This derivation clearly shows that once the multi-linear SVD is computed, coil compression can be achieved via a single tensor-matrix multiplication.

In this study, bSSFP datasets were Fourier transformed in the fully-sampled readout dimension prior to coil compression. A higher-order SVD (HOSVD) algorithm proposed in \cite{de2000multilinear} was used. The learned unitary matrix in the coil dimension was then used to map $D$ original coils in the undersampled dataset onto $D'$ virtual coils. For comparison, software-based compression was also performed via GCC \cite{Zhang:2013df}. Since disjoint sampling patterns are used here, GCC was performed independently for each bSSFP acquisition. MLCC and GCC were both performed over a window of 5 cross-sections in the readout dimension.

All coil-compression algorithms were executed in MATLAB (MathWorks, MA). The implementation of MLCC utilized the TensorLab package \cite{tensorlab3.0}. The MLCC algorithm is available for general use at: \url{http://github.com/icon-lab/mrirecon}.

\subsection*{Simulations}
Balanced SSFP acquisitions of a brain phantom with 0.5 mm isotropic resolution were simulated (http://www.bic.mni.mcgill.ca/brainweb). Signal levels for each tissue were calculated based on \Cref{eq:sind}. The following set of tissue parameters were assumed: T$_1$/T$_2$ of 3000/1000 ms for cerebro-spinal fluid (CSF), 1200/250 ms for blood, 1000/80 ms for white matter, 1300/110 ms for gray matter, 1400/30 ms for muscle, and 370/130 ms for fat \cite{Ilicak:FJpKoYYb}. The sequence parameters were $\alpha=60\degree$ (flip angle), TR/TE=10.0/5.0 ms, and $\Delta\phi $=2$\pi \frac{[0:1:(N-1)]}{N}$. The simulations were based on a realistic distribution of main-field inhomogeneity yielding 0$\pm$62 Hz (mean$\pm$std across volume) off-resonance. An array of 8 coils in a circular configuration within each 2D cross-section was assumed. Multi-coil images were simulated by multiplying each phase-cycled bSSFP image with analytically-derived coil sensitivities \cite{Allison:2013ej}. 

Simulated acquisitions were each undersampled by a factor of R. Here disjoint sampling based on variable-density random pattterns was used \cite{Cukur:2015ic}, which we previously observed to outperform uniform-density and Poisson-disc sampling in phase-cycled bSSFP imaging \cite{Ilicak:FJpKoYYb}. Isotropic acceleration was implemented in two phase-encode dimensions based on a polynomial sampling density function \cite{MikiCS}. A central k-space region was fully sampled for calibration of the interpolation kernel. 

Undersampled data were then reconstructed via ZF, SPIRiT, PE-SSFP and ReCat. Reconstruction quality was assessed with the peak signal-to-noise ratio (PSNR) metric. To prevent bias due to differences in image scale, the 98$^{th}$ percentile of intensity values were mapped onto the $[0, 1]$ range. To prevent bias from background regions void of tissues, a tissue mask was generated for each cross-section by simple thresholding. Images were masked prior to PSNR calculation. The reference image was taken as the Fourier reconstructions of fully-sampled acquisitions at N=8. All metrics were pooled across the central cross-sections of 10 different simulated phantoms. 

To optimize reconstruction and sampling parameters, undersampled data were processed with varying $\beta \in (0,0.2]$, $\lambda \in (0,0.03]$, and radius of calibration region $\in$ [4\%,20\%] of the maximum spatial frequency. Varying p-norm combination parameters were also considered across coils ($p_{coils}$) and across phase-cycles ($p_{acq}$) $\in$ [1,5]. The quality of reconstructions was assessed via the peak signal-to-noise ratio (PSNR) metric, which is a logarithmic measure inversely proportional to the mean squared error (MSE) between a reconstructed image and a reference image. Representative results for N=4, D=8 are shown in Sup.~Fig. S1 and Sup.~Fig. S2. The optimized parameters for ReCat were $\beta=0.05$, $\lambda=0.018$, a calibration region of radius 13\%, $p_{coils}=2$, and $p_{acq}=4$. These parameters also enabled SPIRiT and PE-SSFP to achieve more than 99.0\% of their optimal performance. Therefore, this parameter set was prescribed for all reconstructions thereafter.

To validate the optimization algorithm in ReCat, two different implementations were considered based on LSQR and projection onto convex sets (POCS) methods \cite{Lustig:2010hs}. Reconstructions were obtained for the same set of parameters including number of iterations. Bivariate Gaussian noise was added to simulated acquisitions to attain SNR=20, where SNR was taken as the ratio of total power in k-space data to the power of noise samples. Representative images via LSQR and POCS methods are shown in Sup.~Fig.~S3 for N=4, D=8, and R=12. LSQR maintains lower reconstruction errors, with 0.6 dB higher PSNR than POCS, implying improved convergence properties.

To test robustness against variability in tissue and sequence parameters, extended simulations over equispaced cross-sections of a single subject were performed for varying T$_1$/T$_2$ ratios, flip angles, TRs (with TE=TR/2), and SNR levels. The following range of parameters were considered: (-20\%, 0\%, 20\%) deviation in T$_1$/T$_2$ ratios, $\alpha=(30\degree, 60\degree, 90\degree)$, TR=(5 ms, 10 ms, 15 ms), and SNR levels in [10, 30].

\subsection*{In Vivo Experiments}
In vivo bSSFP acquisitions of the brain were performed on a 3 T Siemens Magnetom scanner (with 45 mT/m maximum gradient strength and 200 T/m/s). A 3D Cartesian bSSFP sequence was prescribed with a flip angle of 30$\degree$, a TR/TE of 8.08 ms/4.04 ms, a field-of-view (FOV) of 218 mm, a resolution of 0.85$\times$0.85$\times$0.85 mm$^3$, elliptical scanning, and N=8 separate acquisitions with $\Delta\phi$ spanning $[0, 2\pi)$ in equispaced intervals. A readout bandwidth of 199 Hz/pixel was used to increase acquisition SNR and thereby improve reconstruction performance at high R. Standard volumetric shimming was performed. Prior to each phase-cycled acquisition, a start-up segment with 10 dummy TRs was used to dampen transient signal oscillations. Each fully-sampled acquisition lasted 2 min 37 s, yielding a total scan time of nearly 21 min. The acquisitions for each subject were collected sequentially, without delay in a single session. Two separate experiments were conducted, the first one using a 12-channel receive-only head coil that was hardware-compressed to 4 output channels, and the second one using a 32-channel receive-only head coil for demonstration of MLCC. The number of participants were 8 for the first experiment and 6 for the second experiment. All participants gave written informed consent, and the imaging protocols were approved by the local ethics committee at Bilkent University.

In vivo bSSFP acquisitions of the brain were variable-density undersampled in the two phase-encode dimensions retrospectively to attain R $\in$ [4, 16] (where R is the acceleration rate with respect to a fully-sampled acquisition). ZF, SPIRiT, PE-SSFP and ReCat were subsequently performed. The following subsets of acquisitions were selected for varying N: $\Delta\phi $=2$\pi \frac{[0:1:(N-1)]}{N}$ for N=2, 4 and 8.

To compare coil-compression techniques, 32-channel acquisitions were reduced to 6 virtual coils that capture nearly 78\% of the total variance in data. GCC and MLCC compressions were separately obtained. ZF, SPIRiT, PE-SSFP and ReCat reconstructions were performed on the compressed datasets. To examine the effect of $D'$ on compression  performance, GCC and MLCC were performed for varying number of virtual coils $D'$=[3, 8]. Separate ReCat reconstructions were computed for each $D'$ while R=[4, 16] and N=[2, 8]. To examine the effect of MLCC on information captured by virtual coils, the variances explained by MLCC and GCC were compared at each $D'$ value. To assess the amount of shared information among phase-cycles in compressed images, Pearson's correlation coefficient was calculated between each pair of phase-cycles. 

To examine image quality, PSNR was measured across the central cross-section in the readout dimension for each subject. Significant differences among reconstructions were assessed with nonparametric Wilcoxon signed-rank tests. Similar to simulation analyses, images were masked to select tissue regions prior to measurements. The reference image was taken as the combined Fourier reconstruction of fully-sampled, uncompressed acquisitions with N=8.

\clearpage
\section*{Results}
\subsection*{Simulations}
\begin{figure}[t]
  \begin{center}
    \includegraphics[width=\ohwidth]{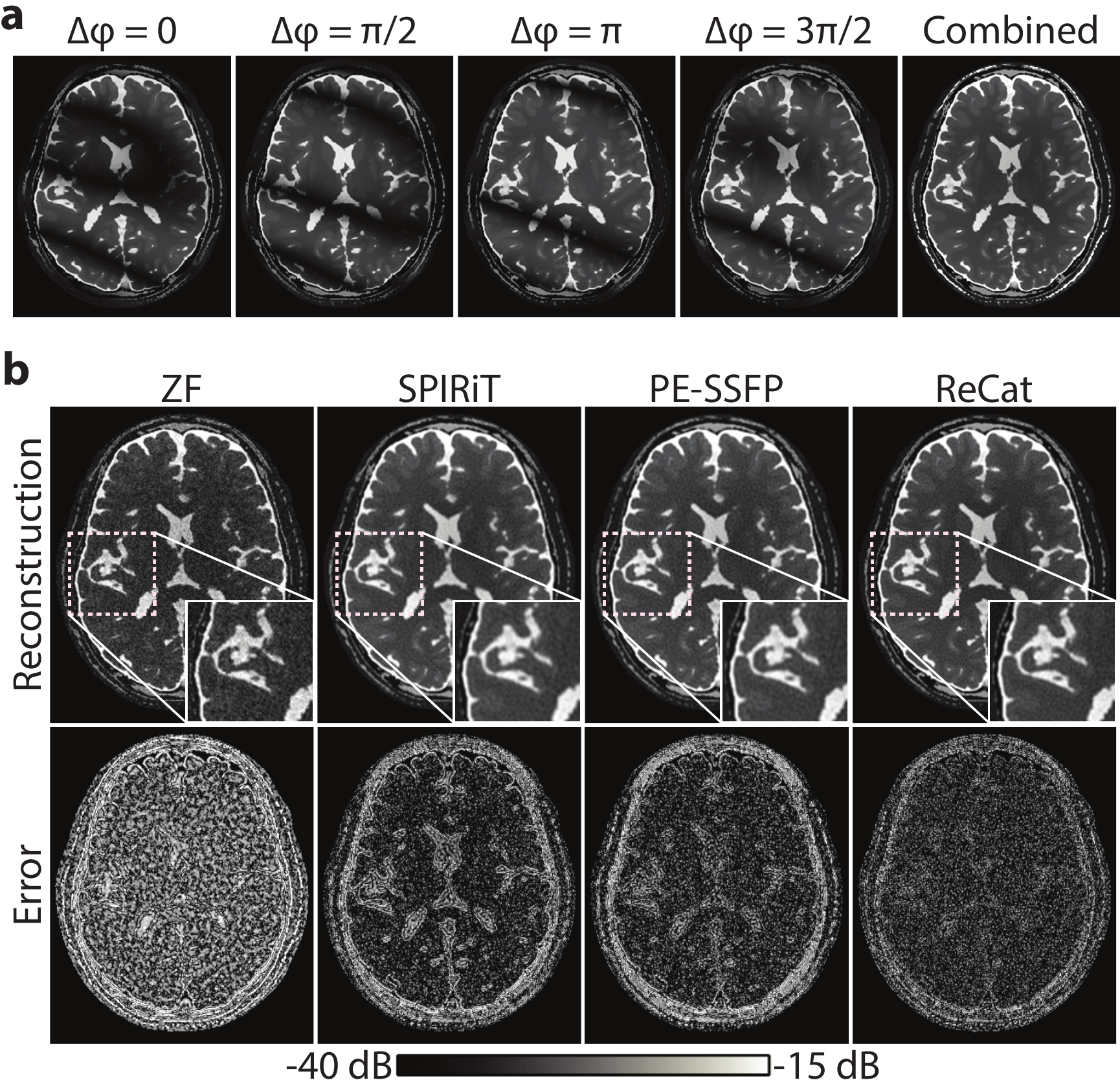}
    \caption{Phase-cycled bSSFP acquisitions of a brain phantom were simulated for N=4. \textbf{(a)} Phase-cycle images and the p-norm combined reference image are shown. \textbf{(b)} Representative reconstructions at  D=8, R=8 are shown for ZF, SPIRiT, PE-SSFP and ReCat (top row). Zoomed-in portions of the images are depicted in small display windows. ReCat depicts detailed tissue structure with greater acuity compared to other methods. Error maps relative to fully-sampled acquisitions are displayed in logarithmic scale (bottom row; see colorbar). ReCat visibly reduces reconstruction errors compared to alternative methods. For this cross-section, ReCat achieves 30.6 dB PSNR, whereas SPIRiT and PE-SSFP yield 29.6 dB and 29.4 dB, respectively. 
    }
    \label{fig:phantom}
  \end{center}
\end{figure}

\begin{figure}[t]
	\begin{center}
		\includegraphics[width=\swidth]{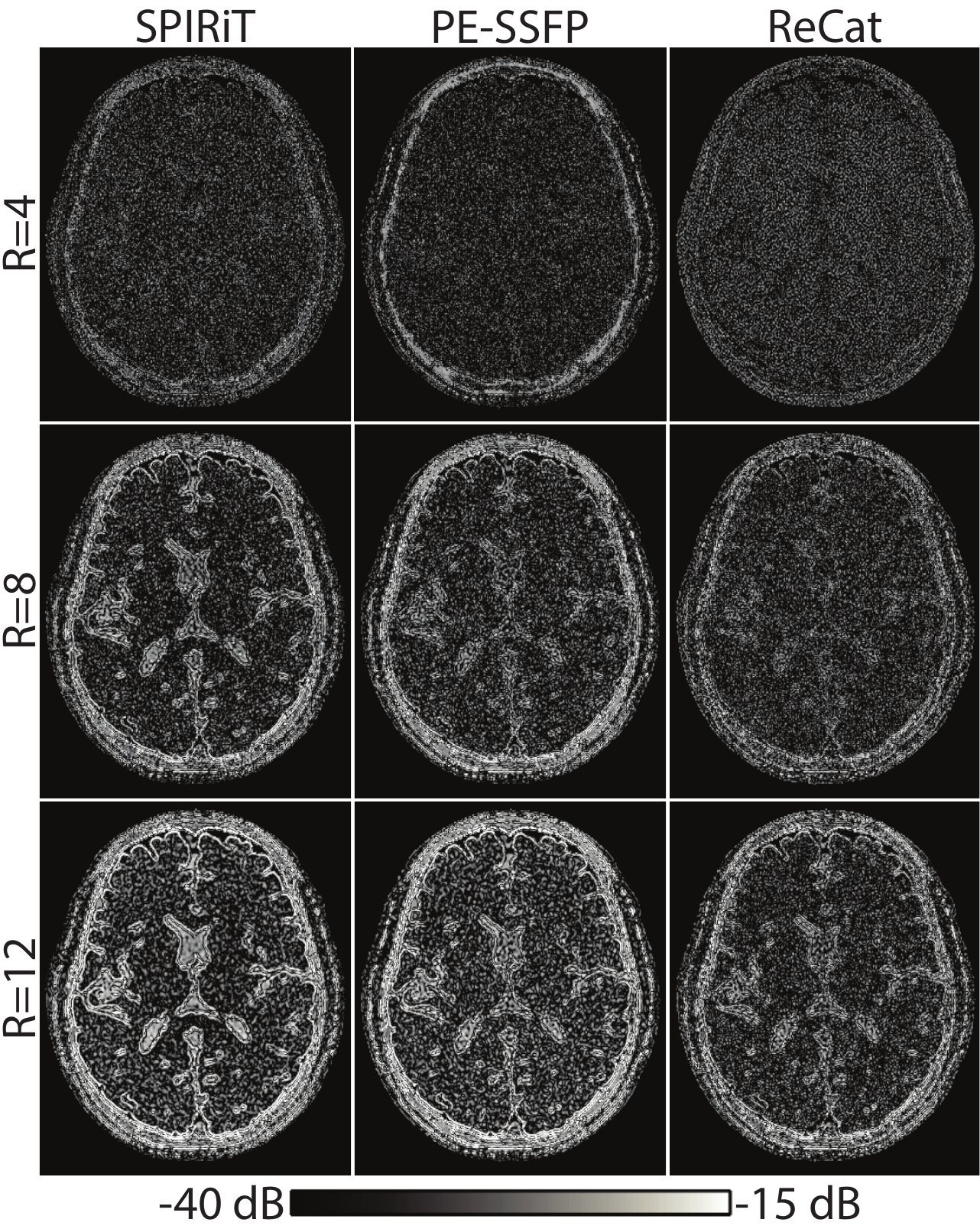}
		\caption{SPIRiT, PE-SSFP and ReCat reconstructions of the simulated brain phantom were performed at N=4 and D=8. Error maps are shown for R=4, 8 and 12. ReCat outperforms SPIRiT and PE-SSFP for R$\textgreater$4, and the level of error reduction increases towards higher R.
		}
		\label{fig:undersampling}
	\end{center}
\end{figure}

ReCat was first demonstrated on bSSFP acquisitions of a numerical brain phantom with D=8. ZF, SPIRiT, PE-SSFP and ReCat reconstructions and error maps are shown in \Cref{fig:phantom}. Error maps for varying acceleration factors R=$\{4,8,12\}$ are shown in Fig.~\ref{fig:undersampling}. SPIRiT that independently processes separate acquisitions and PE-SSFP that independently processes separate coils suffer from broad errors at high-spatial frequencies. In comparison, ReCat achieves visibly reduced reconstruction error and enhanced tissue depiction, particularly for R$\textgreater$4.

Quantitative assessments regarding ReCat and alternative reconstructions are listed in \Cref{tab:phantom} for N=2-8 and R=4-16. ReCat yields higher PSNR values compared to SPIRiT and PE-SSFP at all N and R, except for two cases R=4, N=8 and R=4, N=4 where the techniques perform similarly. On average, ReCat improves PSNR by 2.0$\pm$1.0 dB over SPIRiT, and by 2.0$\pm$0.5 dB over PE-SSFP (mean$\pm$std across subjects; average for N=2-8, R=8-16).

Extended simulations presented in Sup. Tables~S1-S4 indicate that the ReCat provides similar performance improvements over alternative reconstructions broadly across varying noise levels (SNR=10-30), TRs (5-15 ms), flip angles ($30\degree$-$90\degree$), and T$_1$/T$_2$ ratios (-20\% to 20\%). These results suggest that ReCat enhances image quality and improves artifact suppression compared to reconstructions that ignore correlated information across coils or acquisitions. 

\subsection*{In Vivo Experiments}
\begin{figure}[t]
  \begin{center}
    \includegraphics[width=\ohwidth]{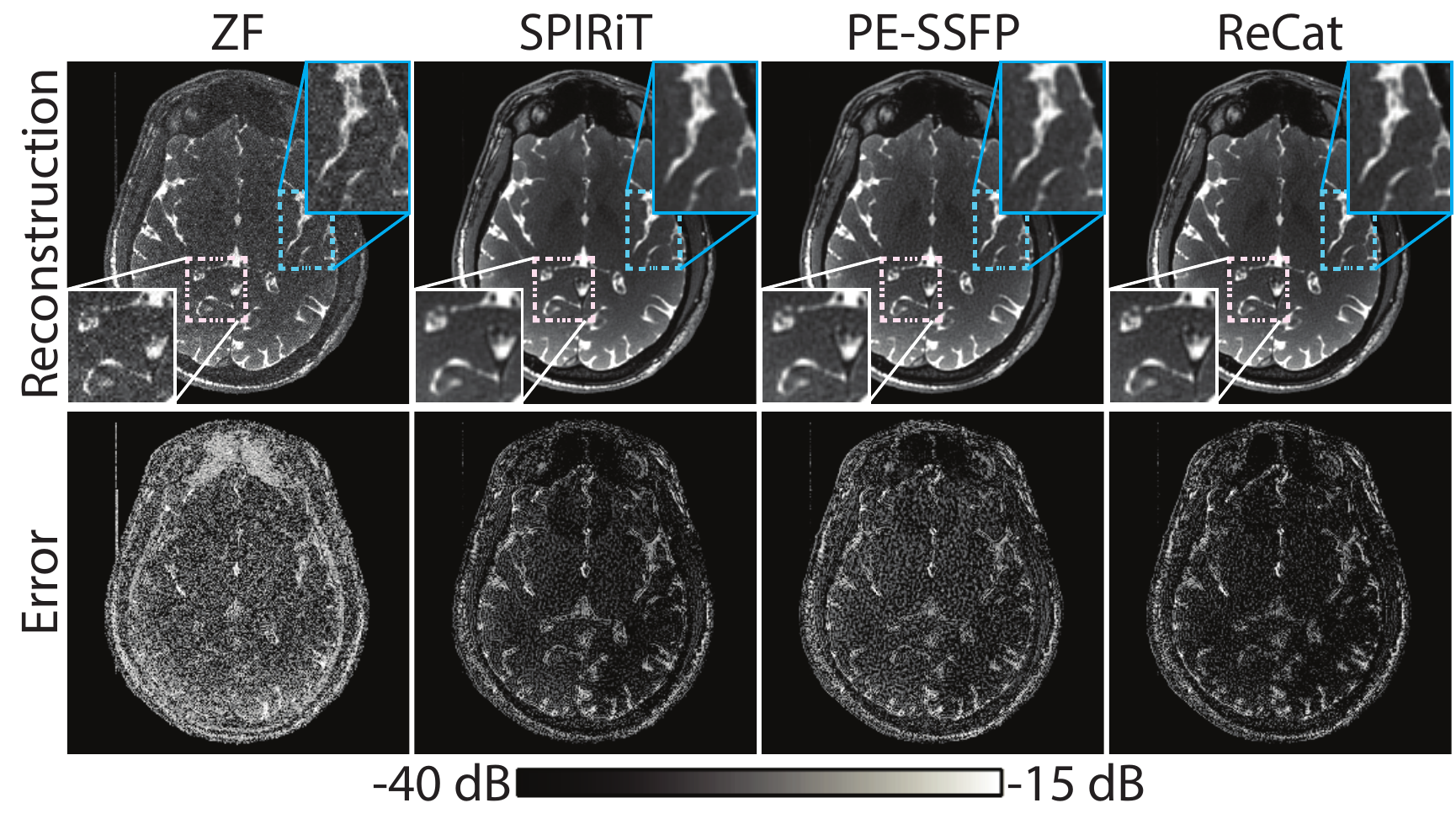}
    \caption{In vivo bSSFP acquisitions of the brain were performed for N=4, D=32. Representative reconstructions at R=8 are shown for ZF, SPIRiT, PE-SSFP and ReCat (top row). Error maps relative to fully-sampled acquisitions are displayed in logarithmic scale (bottom row; see colorbar). ReCat reduces reconstruction error and suppresses artifacts compared to other approaches, and achieves 34.1 dB PSNR; while SPIRiT and PE-SSFP yield to 33.7 dB and 32.4 dB, respectively. (See also Sup. Fig. S4.)
    }
    \label{fig:invivo}
  \end{center}
\end{figure}

Following simulations, the potential of ReCat for accelerated in vivo bSSFP imaging was examined in the brain. Representative images from ZF, SPIRiT, PE-SSFP and ReCat are displayed for D=12 in Sup.~Fig.~S4, and for D=32 in Fig.~\ref{fig:invivo}. For D=12, ZF and SPIRiT suffer from relatively high levels of residual aliasing and noise interference compared to PE-SSFP and ReCat. While ReCat maintains the lowest reconstruction error, PE-SSFP and ReCat images are visually similar with detailed depiction of tissue structure even at high R. For D=32, ReCat again yields high-quality images, and in this case ReCat images appear sharper than PE-SSFP images. As opposed to PE-SSFP that jointly processes acquisitions, ReCat leverages additional information across coils. Thus, as D increases relative to N, performance improvements that ReCat provides over PE-SSFP might become more prominent.

Quantitative assessments of in vivo reconstructions are listed in Table~\ref{tab:invivo} for D=12, N=2-8 and R=4-16. ReCat achieves higher PSNR than SPIRiT for R$\textgreater$4 (p$<$0.05, sign-rank test), and higher PSNR than PE-SSFP for all N and R (p$<$0.05). On average, ReCat improves PSNR by 1.1$\pm$1.0 dB over SPIRiT, and by 0.9$\pm$0.3 over PE-SSFP (mean$\pm$std across subjects; average for N=2-8, R=8-16). 

\begin{figure}[t]
	\begin{center}
		\includegraphics[width=\ohwidth]{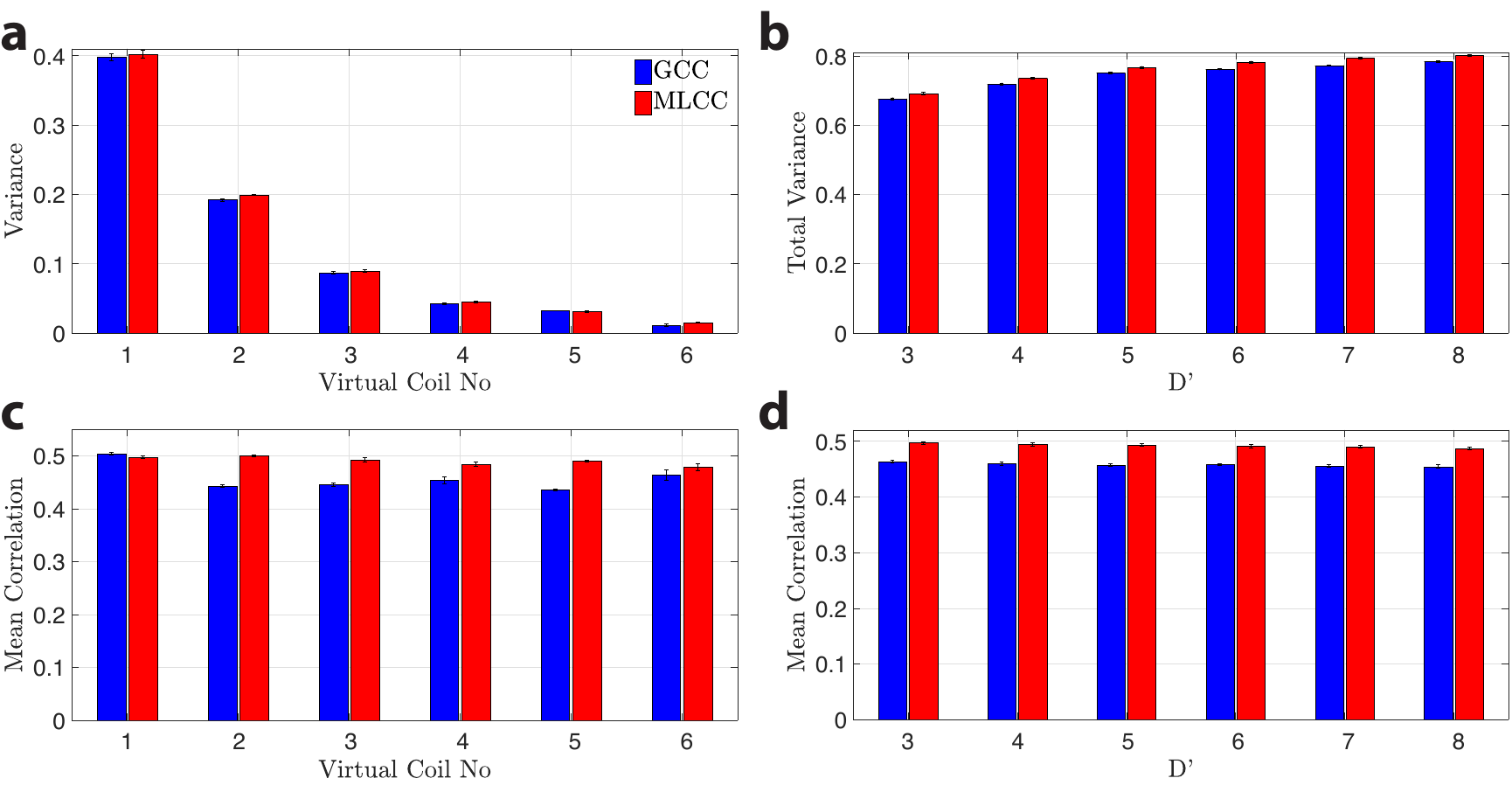}
		\caption{In vivo bSSFP acquisitions of the brain were performed with N=8, D=32, R=8. Coil compression was performed via GCC and MLCC. The bar plots show the mean and standard error across 5 cross-sections. \textbf{(a)} For each virtual coil, the average proportion of variance captured across phase-cycles is plotted when $D'=6$. \textbf{(b)} For varying $D'=[3,8]$, the total proportion of variance captured by all virtual coils is shown. \textbf{(c)} For each virtual coil, the average correlation coefficient of virtual coils across phase-cycles is plotted when $D'=6$. \textbf{(d)} For varying $D'=[3,8]$, the average of all pair-wise correlations of virtual coils is shown. 	
		}
		\label{fig:ccinfo}
	\end{center}
\end{figure}

Next, the proposed coil compression –MLCC– was demonstrated on multi-coil data with D=32. Figure~\ref{fig:ccinfo} displays the proportion of variance that is captured by $D'$=6 virtual coils, and the average correlation coefficient between pairs of virtual coil images for a representative subject. MLCC slightly improves variance explained in virtual coils compared to GCC. Furthermore, it increases the amount of shared information across acquisitions captured in coil-compressed data. This can be confirmed visually by virtual coils shown in Sup.~Fig.~S5. While coil sensitivities based on GCC vary substantially among acquisitions, MLCC yields more consistent coil sensitivities. Note that each acquisition in MLCC-based coils still shows intensity modulation due to bSSFP profiles. These results are valid in each individual subject. Because ReCat leverages an interpolation kernel to synthesize unacquired data across coils and acquisitions, consistency of virtual coils should enhance interpolation performance. 

\begin{figure}[t]
  \begin{center}
    \includegraphics[width=\ohwidth]{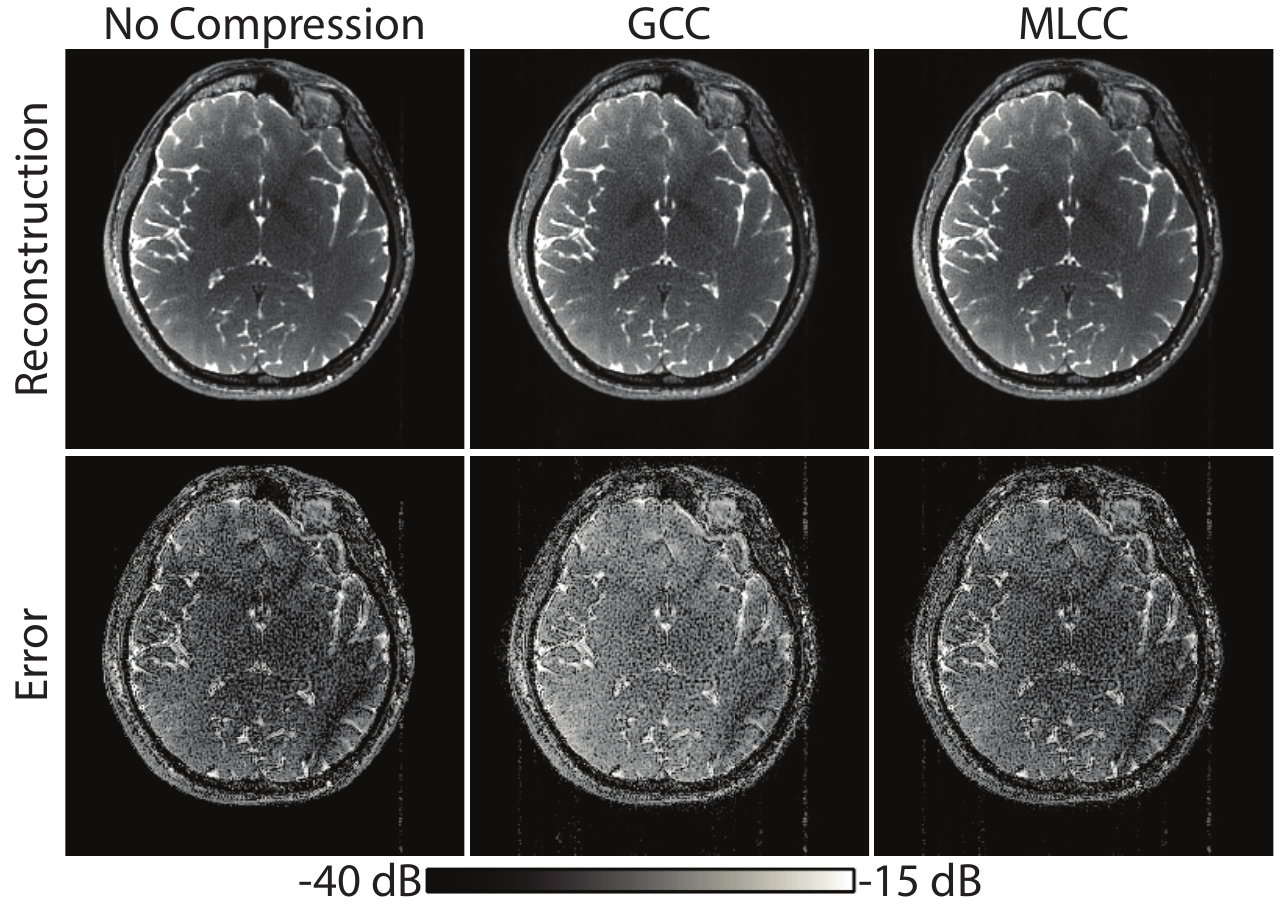}
    \caption{In vivo bSSFP acquisitions of the brain were performed with N=4, D=32. Multi-coil data were compressed to 6 virtual coils (capturing nearly 78\% of the data variance) via GCC and MLCC. Representative ReCat reconstructions are shown at R=8 (top row) along with error maps relative to fully-sampled, uncompressed acquisitions (bottom row). MLCC outperforms GCC, and it produces a reconstruction that more closely resembles the reference no-compression case. Compared to 30.7 dB PSNR in the no-compression case, MLCC yields 30.5 dB PSNR while GCC yields merely 28.7 dB PSNR.
    }
    \label{fig:coil_compression}
  \end{center}
\end{figure}

ReCat reconstructions and respective error maps following GCC and MLCC with $D'$=6 virtual coils are displayed in Fig.~\ref{fig:coil_compression}. For SPIRiT, PE-SSFP and ReCat, MLCC enables substantially reduced errors compared to GCC, as it increases the amount of information in virtual coils that is shared across multiple acquisitions. Quantitative assessments of coil-compressed ReCat reconstructions are listed in Table~\ref{tab:coil_compression} for N=2-8, R=4-16, and $D'$=6. A comprehensive list of measurement for various reconstruction methods is in Sup. Table~S5. For ZF, MLCC and GCC show no significant differences since they account for similar proportion of variance in coil data. For SPIRiT, PE-SSFP and ReCat, MLCC yields higher PSNR than GCC for all N and R (p$\textless$0.05, sign-rank test). On average, MLCC improves PSNR by 0.8$\pm$0.6 dB over GCC for ReCat (mean$\pm$std across subjects; average for N=2-8, R=4-16). 

Differences in PSNR of ReCat images obtained after MLCC and GCC are plotted in Sup.~Fig.~S6 for varying $D'$=[3, 8] in a representative subject. For $D'\textgreater$4, MLCC consistently improves PSNR over GCC regardless of R or N. Taken together, these results suggest that the proposed framework enables scan-efficient phase-cycled bSSFP imaging at high R with improved image quality due to the tensor-based reconstruction and coil compression.

\clearpage
\section*{Discussion}
Several lines of work have produced successful approaches to suppress banding artifacts in bSSFP imaging. Proposed methods for alleviating sensitivity to field inhomogeneity include modification of magnetization profiles \cite{wideband,Benkert:2014hp,Sun:2015ct}, advanced shimming \cite{Lee:2009hq}, and phase-cycled imaging \cite{Bangerter04}. Compared to methods that require pulse-sequence modification, phase-cycled bSSFP with its ease of implementation has remained a popular choice albeit at the expense of prolonged scan times. 

To improve scan efficiency in phase-cycled bSSFP, we recently proposed a profile-encoding approach (PE-SSFP) that jointly reconstructs multi-acquisition data \cite{Ilicak:FJpKoYYb}. PE-SSFP was demonstrated to outperform both independent CS \cite{Cukur:2015ic} and multi-coil PI reconstructions \cite{Wang:2015hh} of individual acquisitions. Since it utilizes correlated structural information across acquisitions, PE-SSFP could maintain high image quality up to R=6-8. However, it remains suboptimal since data from each coil were treated independently. 

In this study, we proposed an improved acceleration framework, ReCat, that linearly synthesizes unacquired data using a tensor-interpolation kernel over coil and acquisition dimensions. We further proposed a tensor-based coil compression, MLCC, that jointly processes acquisitions to produce consistent sets of virtual coils. MLCC improves ReCat by enabling more optimal use of shared information across acquisitions, particularly for disjoint sampling. With this enhanced framework, detailed tissue depiction was maintained up to R=16 and N=8. Thus nearly two-fold increase in scan efficiency was attained while prescribing a large number of acquisitions that effectively suppress banding artifacts. Compared to SPIRiT and PE-SSFP, ReCat yields significantly higher PSNR for simulated phantom and in vivo brain datasets. ReCat also improves image sharpness over SPIRiT and noise and artifact suppression over PE-SSFP. Future studies on a patient population are warranted to assess whether ReCat improves diagnostic quality for radiological evaluations. 

ReCat outperforms both SPIRiT and PE-SSFP for relatively high acceleration factors, but we observed that SPIRiT yields higher PSNR for R$\leq$4. In theory, the higher-dimensional ReCat kernel should yield equal or better performance than the SPIRiT kernel that only captures the coil dimension. In practice, however, the fidelity of kernel estimates can decrease with increasing dimensionality. During recovery of heavily undersampled datasets, the ReCat kernel captures additional information about bSSFP profiles to boost reconstruction performance. Yet for densely sampled datasets with R$\leq$4 and a large number of coils, the benefit of bSSFP-profile information is naturally more limited and can be outweighed by performance losses due to decreased kernel fidelity.

ReCat is an acceleration framework proposed primarily for phase-cycled bSSFP imaging. The bSSFP signal model reveals that each acquisition performs spatial encoding via a respective bSSFP profile, analogous to spatial encoding via coil sensitivities. Here, we showed that the tensor-interpolation kernel in ReCat captures this encoding information from calibration data, and outperforms a kernel across coils or a kernel across acquisitions. Note that calibration-free frameworks were recently proposed for sparse recovery via low-rank structured matrix completion \cite{Shin:2013bl, Haldar:2014ei, Jin:2016cz}. These frameworks can offer improved performance in cases where calibration data are scarce or accuracy of kernel estimates is limited. In particular, the annihilating filter-based low rank Hankel matrix approach (ALOHA) uses efficient implementations of low-rank constraints in transform domains to unify PI and CS reconstructions. These improvements can help further reduce residual aliasing and noise interference in reconstruction of bSSFP datasets. That said, a fair comparison among frameworks requires implementations based on similar types of regularization terms. Currently, ReCat is cast as a linear problem with $\ell_2$-regularization on reconstructed data. We plan to incorporate $\ell_1$-norm, total variation, and low-rank constraints in ReCat to perform comprehensive evaluations in future studies. 

Several technical limitations might be further addressed to improve the proposed framework. First, while scan acceleration partly alleviates motion sensitivity, separate phase cycles are acquired sequentially in ReCat. If significant motion occurs in between the collection of central k-space data for separate phase cycles, joint reconstruction might be impaired due to spatial displacement. To address this issue, motion correction could be incorporated into the reconstructions \cite{Aksoy:2012gi}. Motion can also alter the spatial distribution of field-inhomogeneity-induced phase across multiple acquisitions. ReCat can estimate interpolation kernels that take into account alterations in the encoded bSSFP profiles. However, since these profiles may no longer correspond to phase-cycles equispaced in [0, $2\pi$), higher noise amplification may be observed in the reconstructions. Lastly, for very high R approaching 16, the preparation time needed to reach steady state for each phase-cycled acquisition can become comparable to the acquisition time itself. In such cases, a prepatory segment of ten dummy excitations may prove insufficient in suppressing transient oscillations. To better dampen oscillations, the prepatory segment can be prolonged and advanced preparations based on gradually-ramped RF flip angles might be used \cite{Hargreaves:2001vd}. Still, scan efficiency considerations can impose an upper limit on the achievable acceleration factors. 

ReCat produces images for each individual coil and acquisition separately. Here, these images were combined across both dimensions with the p-norm method to attain a favorable compromise between signal homogeneity and SNR efficiency. A simple sum-of-squares combination (p=2) for coils may lead to suboptimal efficiency at higher noise levels. In such cases, an SNR-optimal linear combination could be performed instead \cite{Cukur:2008ht}. The homogeneity of the p-norm combination (p=4) for bSSFP may also degrade when imaging at high field strengths. To improve homogeneity, analytical methods can be used to better separate the signal components due to tissue parameters and those due to off-resonance \cite{Xiang:2014gc,Bjork:2014hm}. Other ReCat parameters including regularization weights and calibration area size were optimized on simulated phantoms, and then used to reconstruct all datasets in this study. With these parameters, ReCat maintains similar performance improvements across a wide range of sequence and tissue parameters. When larger deviations in scan protocols are expected, it might be preferable to reoptimize ReCat parameters on training data acquired with each unique protocol. Here, a long-TR bSSFP sequence with low readout bandwidth was used to improve reconstruction performance at high R. Similar acquisition time and image quality can also be maintained via a short-TR sequence with higher readout bandwidth and lower R. While this short-TR sequence may further decrease sensitivity to field inhomogeneity, the long-TR sequence can allow for multi-echo bSSFP acquisitions \cite{Reeder:2005gu} and yield improved arterial-venous blood contrast for angiographic applications \cite{Cukur:2011iy}.

The MLCC method proposed here uses the HOSVD algorithm to decompose the multi-coil, multi-acquisition data tensor. Although rarely encountered, low-rank approximations based on HOSVD can recover local optima \cite{de2000best}. In such cases, optimization-based algorithms can be used at the expense of increased computational load \cite{cichocki2015tensor}. For the datasets considered here, no significant differences were observed between HOSVD and optimization-based SVD solutions. Thus HOSVD was preferred for its computational efficiency. In addition to multiple acquisitions, the proposed MLCC method also leverages shared information across multiple cross-sections. Here high quality compression was obtained with MLCC on five cross-sections. This strategy might be suboptimal in cases with substantial, nonsmooth changes in coil sensitivity or tissue structure through cross-sections. The optimal number of cross-sections for MLCC will be application-specific, and it warrants further investigation. 

In summary, ReCat significantly improves scan efficiency of bSSFP imaging while maintaining reliability against field inhomogeneity. By leveraging shared information across both acquisitions and coils, it achieves enhanced image quality compared to conventional PI and CS methods. The computational complexity of the joint reconstruction is effectively addressed via the MLCC method. To optimize image quality, MLCC produces a consistent set of virtual coils across separate acquisitions. The possibility of accelerated brain imaging via multiple phase-cycled bSSFP acquisitions was demonstrated in the current study. Yet, the suggested benefits of ReCat are expected to generalize to many multi-acquisition bSSFP applications including peripheral angiography \cite{Cukur:2011iy}, magnetization transfer imaging \cite{Bieri:2008fl} and fat/water separation \cite{MultipleTR}. Moreover, ReCat and MLCC can be adapted to other multiple-acquisition applications, such as multi-echo fat/water separation \cite{ReederIDEAL}, parametric mapping \cite{Doneva:2010fe, Lee:2016jh}, or dynamic imaging \cite{Jung:2009ir} where there is substantial shared structural information across acquisitions.

\begin{table}[t]
	
	\centering
	\footnotesize
	\setlength{\tabcolsep}{4pt}
	\caption{Measurements on Simulated Phantoms}
	\label{tab:phantom} 
	\begin{threeparttable}
		\begin{tabular*}{\mwidth}{@{\extracolsep{\fill}}l|l|cccc} 
			\multicolumn{6}{c}{\textbf{}} \\[.5ex]
			\hline 
			\multicolumn{2}{c|}{\textbf{}} \T & \textbf{R=4} & \textbf{R=8} & \textbf{R=12} & \textbf{R=16}\\ 
			\hline \hline \T 
			
			\multirow{4}{*}{\textbf{N=8}}
			& {\textbf{ZF}}  \T & 23.5$\pm$0.1 & 22.5$\pm$0.2 & 20.2$\pm$0.3 & 15.8$\pm$0.2\\ 
			\cline{2-6}
			& {\textbf{SPIRiT}} \T & 33.6$\pm$0.3 & 29.1$\pm$0.5 & 26.3$\pm$0.6 & 24.4$\pm$0.7\\ 
			\cline{2-6}
			& {\textbf{PE-SSFP}} \T & 34.3$\pm$0.3 & 30.8$\pm$0.3 & 28.0$\pm$0.5 & 26.1$\pm$0.6\\ 
			\cline{2-6}
			& {\textbf{ReCat}} \T & 33.5$\pm$0.3 & 32.0$\pm$0.2 & 29.6$\pm$0.4 & 27.6$\pm$0.5\\ 
			\hline \hline \T
			
			\multirow{4}{*}{\textbf{N=4}}
			& {\textbf{ZF}} \T & 24.0$\pm$0.1 & 22.2$\pm$0.3 & 18.6$\pm$0.3 & 15.0$\pm$0.3\\ 
			\cline{2-6}
			& {\textbf{SPIRiT}} \T & 32.4$\pm$0.2 & 28.6$\pm$0.4 & 26.1$\pm$0.6 & 24.3$\pm$0.7\\ 
			\cline{2-6}
			& {\textbf{PE-SSFP}} \T & 32.4$\pm$0.3 & 28.6$\pm$0.4 & 26.2$\pm$0.6 & 24.5$\pm$0.7\\ 
			\cline{2-6}
			& {\textbf{ReCat}} \T & 32.5$\pm$0.2 & 30.5$\pm$0.2 & 28.3$\pm$0.4 & 26.4$\pm$0.5\\ 
			\hline \hline \T
			
			\multirow{4}{*}{\textbf{N=2}}
			& {\textbf{ZF}} \T & 24.1$\pm$0.2 & 20.6$\pm$0.3 & 16.6$\pm$0.3 & 14.5$\pm$0.2\\ 
			\cline{2-6}
			& {\textbf{SPIRiT}} \T & 30.1$\pm$0.2 & 27.6$\pm$0.4 & 25.5$\pm$0.5 & 23.9$\pm$0.6\\ 
			\cline{2-6}
			& {\textbf{PE-SSFP}} \T & 29.0$\pm$0.4 & 25.7$\pm$0.6 & 23.7$\pm$0.7 & 22.5$\pm$0.7\\ 
			\cline{2-6}
			& {\textbf{ReCat}} \T & 30.2$\pm$0.2 & 28.2$\pm$0.3 & 26.3$\pm$0.5 & 24.7$\pm$0.6\\ 
			\hline

		\end{tabular*}
		\begin{tablenotes}
			\item Peak SNR (PSNR) measurements on simulated brain phantoms with $D=8$ and a range of $N$ and $R$. For each reconstruction method, metrics are reported as mean$\pm$std across the central cross-sections of 10 different subjects.
		\end{tablenotes}
	\end{threeparttable}
\end{table}

\begin{table}[t]
	
	\centering
	\footnotesize
	\setlength{\tabcolsep}{4pt}
	\caption{Measurements on In Vivo Data}
	\label{tab:invivo} 
	\begin{threeparttable}
		\begin{tabular*}{\mwidth}{@{\extracolsep{\fill}}l|l|cccc} 
			\multicolumn{6}{c}{\textbf{}} \\[.5ex]
			\hline 
			\multicolumn{2}{c|}{\textbf{}} \T & \textbf{R=4} & \textbf{R=8} & \textbf{R=12} & \textbf{R=16}\\ 
			\hline \hline \T 
			
			\multirow{4}{*}{\textbf{N=8}}
			& {\textbf{ZF}} \T & 28.6$\pm$0.9 & 23.6$\pm$1.2 & 19.1$\pm$1.2 & 14.7$\pm$1.1\\ 
			\cline{2-6}
			& {\textbf{SPIRiT}} \T & 34.8$\pm$1.3 & 29.0$\pm$1.4 & 26.4$\pm$1.4 & 24.7$\pm$1.5\\
			\cline{2-6}
			& {\textbf{PE-SSFP}} \T & 33.6$\pm$1.4 & 29.8$\pm$1.3 & 27.7$\pm$1.2 & 26.1$\pm$1.2\\ 
			\cline{2-6}
			& {\textbf{ReCat}} \T & 34.8$\pm$1.4 & 30.9$\pm$1.2 & 28.6$\pm$1.2 & 27.0$\pm$1.2\\ 
			\hline \hline \T 
			
			\multirow{4}{*}{\textbf{N=4}}
			& {\textbf{ZF}} \T & 25.3$\pm$1.5 & 21.0$\pm$1.3 & 16.9$\pm$1.0 & 13.9$\pm$0.7\\ 
			\cline{2-6}
			& {\textbf{SPIRiT}} \T & 28.3$\pm$2.1 & 26.4$\pm$1.2 & 24.8$\pm$1.0 & 23.6$\pm$1.1\\ 
			\cline{2-6}
			& {\textbf{PE-SSFP}} \T & 27.6$\pm$2.0 & 26.1$\pm$1.4 & 24.8$\pm$1.2 & 23.9$\pm$1.2\\ 
			\cline{2-6}
			& {\textbf{ReCat}} \T & 28.2$\pm$2.3 & 26.9$\pm$1.6 & 25.7$\pm$1.2 & 24.7$\pm$1.1\\ 
			\hline \hline \T 
			
			\multirow{4}{*}{\textbf{N=2}}
			& {\textbf{ZF}} \T & 22.7$\pm$1.7 & 18.7$\pm$1.0 & 15.4$\pm$0.5 & 13.5$\pm$0.5\\ 
			\cline{2-6}
			& {\textbf{SPIRiT}} \T & 24.7$\pm$1.9 & 23.9$\pm$1.4 & 23.0$\pm$1.1 & 22.2$\pm$1.0\\ 
			\cline{2-6}
			& {\textbf{PE-SSFP}} \T & 24.4$\pm$1.8 & 23.4$\pm$1.3 & 22.4$\pm$1.2 & 21.6$\pm$1.1\\ 
			\cline{2-6}
			& {\textbf{ReCat}} \T & 24.7$\pm$2.0 & 24.0$\pm$1.5 & 23.3$\pm$1.3 & 22.5$\pm$1.1\\ 
			\hline

		\end{tabular*}
		\begin{tablenotes}
			\item PSNR measurements on in vivo brain images with $D=12$ and a range of $N$ and $R$. For each reconstruction method, metrics are reported as mean$\pm$std across the central cross-sections of 8 different subjects.	
		\end{tablenotes}
	\end{threeparttable}
\end{table}

\begin{table}[t]
	
	\centering
	\footnotesize
	\setlength{\tabcolsep}{4pt}
	\caption{Measurements on Coil-Compressed In Vivo Data}
	\label{tab:coil_compression} 
	\begin{threeparttable}
		\begin{tabular*}{\mwidth}{@{\extracolsep{\fill}}l|l|cccc} 
			\multicolumn{6}{c}{\textbf{}} \\[.5ex]
			\hline 
			\multicolumn{2}{c|}{\textbf{}} \T & \textbf{R=4} & \textbf{R=8} & \textbf{R=12} & \textbf{R=16}\\ 
			\hline \hline \T  
			
			\multirow{2}{*}{\textbf{N=8}}
			& {\textbf{GCC}} \T & 35.3$\pm$4.8 & 32.5$\pm$4.1 & 30.6$\pm$4.0 & 29.2$\pm$3.7\\
			\cline{2-6} 
			& {\textbf{MLCC}} \T  & 36.9$\pm$5.0 & 33.7$\pm$4.2 & 31.7$\pm$3.9 & 30.0$\pm$3.8\\
			\hline \hline \T

			\multirow{2}{*}{\textbf{N=4}}
			& {\textbf{GCC}} \T & 33.3$\pm$2.2 & 31.4$\pm$2.7 & 29.9$\pm$2.7 & 28.6$\pm$2.6\\ 
			\cline{2-6} 
			& {\textbf{MLCC}} \T & 33.8$\pm$2.2 & 32.2$\pm$2.4 & 30.6$\pm$2.4 & 29.4$\pm$2.3\\
			\hline \hline \T 
			
			\multirow{2}{*}{\textbf{N=2}}
			& {\textbf{GCC}} \T & 29.8$\pm$2.2 & 29.1$\pm$2.3 & 28.1$\pm$2.4 & 27.1$\pm$2.3\\
			\cline{2-6} 
			& {\textbf{MLCC}} \T & 30.1$\pm$2.1 & 29.4$\pm$2.2 & 28.6$\pm$2.2 & 27.7$\pm$2.2\\ 
			\hline
			
		\end{tabular*}
		\begin{tablenotes}
			\item PSNR measurements on in vivo brain images acquired with $D=32$. GCC and MLCC coil compression was performed to attain $D'=6$, followed by a ReCat reconstruction. For different N and R, metrics are reported as mean$\pm$std across the central cross-sections of 6 different subjects. Quantitative coil compression results with other reconstruction techniques are in Sup. Table~S5.		
		\end{tablenotes}
	\end{threeparttable}
\end{table}

\section*{Acknowledgment}
The authors thank L. K. Senel for helpful discussions on the implementation of ReCat.

\clearpage
\bibliography{tolgarefs}

\begin{thebibliography}{10}

\bibitem{SchefflerEur}
Scheffler~K, Lehnhardt~S.
\newblock Principles and applications of balanced {SSFP} techniques.
\newblock Eur Radiol 2003; 13:2409--2418.

\bibitem{Bieri05}
Bieri~O, Markl~M, Scheffler~K.
\newblock Analysis and compensation of eddy currents in balanced {SSFP}.
\newblock Magn\ Reson\ Med 2005; 54:129--137.

\bibitem{Bangerter04}
Bangerter~NK, Hargreaves~BA, Vasanawala~SS, Pauly~JM, Gold~GE, Nishimura~DG.
\newblock Analysis of multiple-acquisition {SSFP}.
\newblock Magn\ Reson\ Med 2004; 51:1038--1047.

\bibitem{Hargreaves:2006bn}
Hargreaves~BA, Bangerter~NK, Shimakawa~A, Vasanawala~SS, Brittain~JH,
  Nishimura~DG.
\newblock {Dual-acquisition phase-sensitive fat-water separation using balanced
  steady-state free precession.}
\newblock Magn Reson Imaging 2006; 24:113--122.

\bibitem{ATRDIXON}
\c{C}ukur~T, Nishimura~DG.
\newblock Fat-water separation with alternating repetition time balanced
  {SSFP}.
\newblock Magn\ Reson\ Med 2008; 60:479--484.

\bibitem{Cukur:2009do}
\c{C}ukur~T, Nishimura~DG.
\newblock {Multiple repetition time balanced steady-state free precession
  imaging.}
\newblock Magn Reson Med 2009; 62:193--204.

\bibitem{Cukur:2007dx}
Cukur~T, Bangerter~NK, Nishimura~DG.
\newblock {Enhanced spectral shaping in steady-state free precession imaging.}
\newblock Magn Reson Med 2007; 58:1216--1223.

\bibitem{Elliott07}
Elliott~AM, Bernstein~MA, Ward~HA, Lane~J, Witte~RJ.
\newblock Nonlinear averaging reconstruction method for phase-cycle {SSFP}.
\newblock Magn\ Reson\ Imaging 2007; 25:359--364.

\bibitem{Quist:2012kx}
Quist~B, Hargreaves~BA, \c{C}ukur~T, Morrell~GR, Gold~GE, Bangerter~NK.
\newblock {Simultaneous fat suppression and band reduction with large-angle
  multiple-acquisition balanced steady-state free precession.}
\newblock Magn Reson Med 2012; 67:1004--1012.

\bibitem{Wang:2015hh}
Wang~Y, Shao~X, Martin~T, Moeller~S, Yacoub~E, Wang~DJJ.
\newblock {Phase-cycled simultaneous multislice balanced SSFP imaging with
  CAIPIRINHA for efficient banding reduction; doi:10.1002/mrm.26076.}
\newblock Magn Reson Med 2015; 0:0--0.

\bibitem{PruessmannSENSE}
Pruessmann~KP, Weiger~M, Scheidegger~MB, Boesiger~P.
\newblock {SENSE}: Sensitivity encoding for fast {MRI}.
\newblock Magn\ Reson\ Med 1999; 42:952--962.

\bibitem{GriswoldGRAPPA}
Griswold~MA, Jakob~PM, Heidemann~RM, Nittka~M, Jellus~V, Jianmin~W, Kiefer~B,
  Haase~A.
\newblock Generalized autocalibrating partially parallel acquisition.
\newblock Magn\ Reson\ Med 2002; 47:1202--1210.

\bibitem{Cukur:2015ic}
Cukur~T.
\newblock {Accelerated Phase-Cycled SSFP Imaging With Compressed Sensing}.
\newblock Medical Imaging, IEEE Transactions on 2015; 34:107--115.

\bibitem{Block}
Block~KT, Uecker~M, Frahm~J.
\newblock Undersampled radial {MRI} with multiple coils: {Iterative} image
  reconstruction using a total variation constraint.
\newblock Magn Reson Med 2007; 57:1086--1098.

\bibitem{MikiCS}
Lustig~M, Donoho~D, Pauly~JM.
\newblock Sparse {MRI}: The application of compressed sensing for rapid {MR}
  imaging.
\newblock Magn\ Reson\ Med 2007; 58:1182--1195.

\bibitem{CSENS}
\c{C}ukur~T, Lustig~M, Nishimura~DG.
\newblock Improving non-contrast-enhanced steady-state free precession
  angiography with compressed sensing.
\newblock Magn\ Reson\ Med 2009; 61:1122--1131.

\bibitem{Ilicak:FJpKoYYb}
Ilicak~E, Senel~LK, Biyik~E, Cukur~T.
\newblock {Profile-encoding reconstruction for multiple-acquisition balanced
  steady-state free precession imaging}.
\newblock Magn Reson Med 2016; pp. n/a--n/a.

\bibitem{Majumdar:2011hj}
Majumdar~A, Ward~RK.
\newblock {Accelerating multi-echo T2 weighted MR imaging: analysis prior
  group-sparse optimization.}
\newblock J. Magn. Reson. 2011; 210:90--97.

\bibitem{Bilgic:2011jv}
Bilgic~B, Goyal~VK, Adalsteinsson~E.
\newblock {Multi-contrast reconstruction with Bayesian compressed sensing.}
\newblock Magn Reson Med 2011; 66:1601--1615.

\bibitem{Jin:2016cz}
Jin~KH, Lee~D, Ye~JC.
\newblock {A General Framework for Compressed Sensing and Parallel MRI Using
  Annihilating Filter Based Low-Rank Hankel Matrix}.
\newblock IEEE Transactions on Computational Imaging 2016; 2:480--495.

\bibitem{Lustig:2010hs}
Lustig~M, Pauly~JM.
\newblock {SPIRiT: Iterative self-consistent parallel imaging reconstruction
  from arbitrary k-space.}
\newblock Magn Reson Med 2010; 64:457--471.

\bibitem{mikisFOV}
Lustig~M, Santos~JM, Pauly~JM.
\newblock A super-{FOV} method for rapid {SSFP} banding artifact reduction.
\newblock { In:} Proceedings of the 13th Annual Meeting of ISMRM, Miami Beach,
  2005.  p. 504.

\bibitem{king2010optimum}
King~SB, Varosi~SM, Duensing~GR.
\newblock Optimum snr data compression in hardware using an eigencoil array.
\newblock Magnetic resonance in medicine 2010; 63:1346--1356.

\bibitem{buehrer2007array}
Buehrer~M, Pruessmann~KP, Boesiger~P, Kozerke~S.
\newblock Array compression for mri with large coil arrays.
\newblock Magnetic Resonance in Medicine 2007; 57:1131--1139.

\bibitem{huang2008software}
Huang~F, Vijayakumar~S, Li~Y, Hertel~S, Duensing~GR.
\newblock A software channel compression technique for faster reconstruction
  with many channels.
\newblock Magnetic resonance imaging 2008; 26:133--141.

\bibitem{Zhang:2013df}
Zhang~T, Pauly~JM, Vasanawala~SS, Lustig~M.
\newblock {Coil compression for accelerated imaging with Cartesian sampling.}
\newblock Magn Reson Med 2013; 69:571--582.

\bibitem{Bjork:2014hm}
Bj{\"o}rk~M, Ingle~RR, Gudmundson~E, Stoica~P, Nishimura~DG, Barral~JK.
\newblock {Parameter estimation approach to banding artifact reduction in
  balanced steady-state free precession.}
\newblock Magn Reson Med 2014; 72:880--892.

\bibitem{Scheffler:2003cu}
Scheffler~K, Lehnhardt~S.
\newblock {Principles and applications of balanced SSFP techniques.}
\newblock Eur Radiol 2003; 13:2409--2418.

\bibitem{Cukur:2008ht}
Cukur~T, Lustig~M, Nishimura~DG.
\newblock {Multiple-profile homogeneous image combination: application to
  phase-cycled SSFP and multicoil imaging.}
\newblock Magn Reson Med 2008; 60:732--738.

\bibitem{Murphy:2012hq}
Murphy~M, Alley~M, Demmel~J, Keutzer~K, Vasanawala~S, Lustig~M.
\newblock {Fast $\ell_1$-SPIRiT compressed sensing parallel imaging MRI:
  scalable parallel implementation and clinically feasible runtime.}
\newblock IEEE Trans Med Imaging 2012; 31:1250--1262.

\bibitem{PNORM}
\c{C}ukur~T, Lustig~M, Nishimura~DG.
\newblock Multiple-profile homogenous image combination: {Application} to
  phase-cycled {SSFP} and multi-coil imaging.
\newblock Magn\ Reson\ Med 2008; 60:732--738.

\bibitem{de2000multilinear}
DeLathauwer~L, DeMoor~B, Vandewalle~J.
\newblock A multilinear singular value decomposition.
\newblock SIAM journal on Matrix Analysis and Applications 2000; 21:1253--1278.

\bibitem{tensorlab3.0}
Vervliet~N, Debals~O, Sorber~L, VanBarel~M, DeLathauwer~L.
\newblock Tensorlab 3.0, Mar. 2016, Available online.

\bibitem{Allison:2013ej}
Allison~MJ, Ramani~S, Fessler~JA.
\newblock {Accelerated regularized estimation of MR coil sensitivities using
  augmented Lagrangian methods.}
\newblock IEEE Trans Med Imaging 2013; 32:556--564.

\bibitem{wideband}
Nayak~KS, Lee~HL, Hargreaves~BA, Hu~BS.
\newblock Wideband {SSFP}: Alternating repetition time balanced steady state
  free precession with increased band spacing.
\newblock Magn\ Reson\ Med 2007; 58:931--938.

\bibitem{Benkert:2014hp}
Benkert~T, Ehses~P, Blaimer~M, Jakob~PM, Breuer~FA.
\newblock {Dynamically phase-cycled radial balanced SSFP imaging for efficient
  banding removal.}
\newblock Magn Reson Med 2015; 73:182--194.

\bibitem{Sun:2015ct}
Sun~H, Fessler~JA, Noll~DC, Nielsen~JF.
\newblock {Balanced SSFP-like steady-state imaging using small-tip fast
  recovery with a spectral prewinding pulse.}
\newblock Magn Reson Med 2016; 75:839--844.

\bibitem{Lee:2009hq}
Lee~J, Lustig~M, Kim~DH, Pauly~JM.
\newblock {Improved shim method based on the minimization of the maximum
  off-resonance frequency for balanced steady-state free precession (bSSFP).}
\newblock Magn Reson Med 2009; 61:1500--1506.

\bibitem{Shin:2013bl}
Shin~PJ, Larson~PEZ, Ohliger~MA, Elad~M, Pauly~JM, Vigneron~DB, Lustig~M.
\newblock {Calibrationless parallel imaging reconstruction based on structured
  low-rank matrix completion.}
\newblock Magn Reson Med 2014; 72:959--970.

\bibitem{Haldar:2014ei}
Haldar~JP.
\newblock {Low-Rank Modeling of Local k-Space Neighborhoods (LORAKS) for
  Constrained MRI}.
\newblock Medical Imaging, IEEE Transactions on 2014; 33:668--681.

\bibitem{Aksoy:2012gi}
Aksoy~M, Forman~C, Straka~M, Cukur~T, Hornegger~J, Bammer~R.
\newblock {Hybrid prospective and retrospective head motion correction to
  mitigate cross-calibration errors.}
\newblock Magn Reson Med 2012; 67:1237--1251.

\bibitem{Hargreaves:2001vd}
Hargreaves~BA, Vasanawala~SS, Pauly~JM, Nishimura~DG.
\newblock {Characterization and reduction of the transient response in
  steady-state MR imaging.}
\newblock Magn Reson Med 2001; 46:149--158.

\bibitem{Xiang:2014gc}
Xiang~QS, Hoff~MN.
\newblock {Banding artifact removal for bSSFP imaging with an elliptical signal
  model.}
\newblock Magn Reson Med 2014; 71:927--933.

\bibitem{Reeder:2005gu}
Reeder~SB, Markl~M, Yu~H, Hellinger~JC, Herfkens~RJ, Pelc~NJ.
\newblock {Cardiac CINE imaging with IDEAL water-fat separation and
  steady-state free precession.}
\newblock J Magn Reson Imaging 2005; 22:44--52.

\bibitem{Cukur:2011iy}
\c{C}ukur~T, Shimakawa~A, Yu~H, Hargreaves~BA, Hu~BS, Nishimura~DG,
  Brittain~JH.
\newblock {Magnetization-prepared IDEAL bSSFP: A flow-independent technique for
  noncontrast-enhanced peripheral angiography.}
\newblock J Magn Reson Imaging 2011; 33:931--939.

\bibitem{de2000best}
DeLathauwer~L, DeMoor~B, Vandewalle~J.
\newblock On the best rank-1 and rank-(r 1, r 2,..., rn) approximation of
  higher-order tensors.
\newblock SIAM Journal on Matrix Analysis and Applications 2000; 21:1324--1342.

\bibitem{cichocki2015tensor}
Cichocki~A, Mandic~D, DeLathauwer~L, Zhou~G, Zhao~Q, Caiafa~C, Phan~HA.
\newblock Tensor decompositions for signal processing applications: From
  two-way to multiway component analysis.
\newblock IEEE Signal Processing Magazine 2015; 32:145--163.

\bibitem{Bieri:2008fl}
Bieri~O, Mamisch~TC, Trattnig~S, Scheffler~K.
\newblock {Optimized balanced steady-state free precession magnetization
  transfer imaging}.
\newblock Magn Reson Med 2008; 60:1261--1266.

\bibitem{MultipleTR}
\c{C}ukur~T, Nishimura~DG.
\newblock Multiple repetition time balanced steady-state free precession
  imaging.
\newblock Magn Reson Med 2009; 62:193--204.

\bibitem{ReederIDEAL}
Reeder~SB, Pineda~AR, Wen~Z, Shimakawa~A, Yu~H, Brittain~JH, Gold~GE,
  Beaulieu~CH, Pelc~NJ.
\newblock Iterative decomposition of water and fat with echo asymmetry and
  least-squares estimation {(IDEAL)}: application with fast spin-echo imaging.
\newblock Magn\ Reson\ Med 2005; 54:636--644.

\bibitem{Doneva:2010fe}
Doneva~M, B{\"o}rnert~P, Eggers~H, Stehning~C, S{\'e}n{\'e}gas~J, Mertins~A.
\newblock {Compressed sensing reconstruction for magnetic resonance parameter
  mapping.}
\newblock Magn Reson Med 2010; 64:1114--1120.

\bibitem{Lee:2016jh}
Lee~D, Jin~KH, Kim~EY, Park~SH, Ye~JC.
\newblock {Acceleration of MR Parameter Mapping Using Annihilating Filter-Based
  Low Rank Hankel Matrix (ALOHA)}.
\newblock Magn Reson Med 2016; 76:1848--1864.

\bibitem{Jung:2009ir}
Jung~H, Sung~K, Nayak~KS, Kim~EY, Ye~JC.
\newblock {k-t FOCUSS: A general compressed sensing framework for high
  resolution dynamic MRI}.
\newblock Magn Reson Med 2009; 61:103--116.

\end{thebibliography}

\clearpage
\listoftables
\clearpage
\listoffigures
\clearpage
\section*{List of Supporting Figures and Tables}
{\textit{Sup. Tab. S1}~~~Effects of $\textrm{SNR}$ Variations\dotfill}
{\newline\textit{Sup. Tab. S2}~~~Effects of $\textrm{TR}$ Variations\dotfill}
{\newline\textit{Sup. Tab. S3}~~~Effects of $\alpha$ Variations\dotfill}
{\newline\textit{Sup. Tab. S4}~~~Effects of $T_1/T_2$ Variations\dotfill}
{\newline\textit{Sup. Tab. S5}~~~Measurements on Coil-Compressed In Vivo Data\dotfill}
{\newline\textit{Sup. Fig. S1}~~~Reconstruction quality was examined as a function of the regularization parameters $\beta$, $\lambda$ and the calibration area size. Results are shown for SPIRiT, PE-SSFP and ReCat methods with $N=4$, $D=8$, $R=8$. \textbf{(a)} PSNR measurements on simulated brain phantoms with varying $\beta$, and fixed calibration area size of 13\% and $\lambda$=0.018. All methods are fairly insensitive to the value of $\beta$ in a broad range. \textbf{(b)} PSNR measurements with varying $\lambda$, and fixed calibration area size of 13\% and $\beta$=0.05. \textbf{(c)} PSNR measurements with varying calibration area size, and fixed $\lambda$=0.018 and $\beta$=0.05. In all cases, SPIRiT and PE-SSFP achieve above 99.0\% of their maximum PSNR at the optimal reconstruction parameters for ReCat.\dotfill}
{\newline\textit{Sup. Fig. S2}~~~Reconstruction quality was examined as a function of the p-norm  parameters $p_{acq}$ and $p_{coils}$, which represent the p-norm values to combine acquisitions and coils, respectively. Results are shown as PSNR measurements for SPIRiT \textbf{(a)}, PE-SSFP \textbf{(b)} and ReCat \textbf{(c)} methods with $N=4$, $D=8$, $R=8$. $p_{acq}=4$ and $p_{coils}=2$ were taken as optimal parameters for ReCat. SPIRiT and PE-SSFP achieve above 99.2\% of their maximum PSNR at the optimal p-norm parameters for ReCat.\dotfill}
{\newline\textit{Sup. Fig. S3}~~~Two different implementations of ReCat were considered based on projection onto convex sets (POCS) and least squares (LSQR) algorithms. Reconstruction parameters for the two methods were independently optimized. Reconstructions (top row) and squared error maps in logarithmic scale (bottom row; see colorbar) are shown for N=4, D=8. Overall, LSQR achieves relatively lower reconstruction errors, with 0.6 dB higher PSNR than the POCS implementation.\dotfill}
{\newline\textit{Sup. Fig. S4}~~~In vivo bSSFP acquisitions of the brain were performed for N=8, D=12. \textbf{(a)} Fully-sampled acquisitions for four sample phase cycles and their p-norm combination are shown. \textbf{(b)} Representative reconstructions at R=8 are shown for ZF, SPIRiT, PE-SSFP and ReCat (top row). Error maps relative to fully-sampled acquisitions are displayed in logarithmic scale (bottom row; see colorbar). ReCat reduces reconstruction error compared to other approaches with 29.3 dB PSNR; while SPIRiT and PE-SSFP yield 27.6 dB and 27.9 dB, respectively. While ReCat and PE-SSFP produce visually similar images, ReCat yields sharper reconstructions compared to SPIRiT.\dotfill}
{\newline\textit{Sup. Fig. S5}~~~Coil-compression was performed on undersampled bSSFP data (N=8, R=4) acquired with D=32 physical coils and compressed to D'=6 virtual coils. ZF reconstructions were obtained for each acquisition and each coil separately. \textbf{(a)} Virtual coil images obtained with GCC for two representative acquisitions, $\Delta \phi = \pi$ (top row), $\pi/4$ (bottom row). Coils are shown in separate columns. \textbf{(b)} Virtual coil images obtained with MLCC for the same acquisitions. While GCC-based coil sensitivities show differences across acquisitions (marked with arrows), MLCC-based sensitivities are highly consistent across acquisitions.\dotfill}
{\newline\textit{Sup. Fig. S6}~~~In vivo bSSFP acquisitions of the brain were performed with D=32. Coil-compression via GCC and MLCC was obtained for varying number of virtual coils $D'$=[3, 8], and ReCat was computed. The data variance captured at each $D'$ value is listed in the horizontal axis. \textbf{(a)} PSNR difference between MLCC and GCC for R=[4, 16] and N=8. \textbf{(b)} PSNR difference between MLCC and GCC for N=[2, 8] and R=8. For D'$\textgreater$4, MLCC improves PSNR over GCC regardless of R or N.\dotfill}

\end{document}